\documentclass[a4paper,fleqn,usenatbib]{mnras}

\usepackage{newtxtext,newtxmath}
\usepackage[T1]{fontenc}
\usepackage{ae,aecompl}
\usepackage{graphicx}	
\usepackage{amsmath}	
\usepackage{amssymb}	
\usepackage{enumitem}
\usepackage{rotating}
\usepackage{pdflscape}	

\newcommand{\gc}{$\gamma$\,Cas}
\newcommand{\te}{{\it TESS}}

\title[\te\ lightcurves of \gc\ stars]{\te\ lightcurves of \gc\ stars\thanks{Based on data obtained with the \te\ mission, whose funding is provided by the NASA Explorer Program, as well as on archival data from {\it SuperWASP, BRITE, ASAS, ASAS-SN} and on supporting {\it TIGRE} spectroscopy.}}

\author[Y. Naz\'e et al.]{Ya\"el~Naz\'e$^1$\thanks{F.R.S.-FNRS Senior Research Associate, email: ynaze@uliege.be}, Gregor Rauw$^1$, and Andrzej Pigulski$^{2}$
\\
$^{1}$Groupe d'Astrophysique des Hautes Energies, STAR, Universit\'e de Li\`ege, Quartier Agora (B5c, Institut d'Astrophysique et de G\'eophysique), \\
All\'ee du 6 Ao\^ut 19c, B-4000 Sart Tilman, Li\`ege, Belgium\\
$^{2}$Instytut Astronomiczny, Uniwersytet Wroc\l{}awski, Kopernika 11, 51-622 Wroc\l{}aw, Poland
}

\pubyear{2020}

\begin{document}
\label{firstpage}
\pagerange{\pageref{firstpage}--\pageref{lastpage}}
\maketitle

\begin{abstract}
  \gc\ stars constitute a subgroup of Be stars showing unusually hard and bright X-ray emission. In search for additional peculiarities, we analyzed the \te\ lightcurves of 15 \gc\ analogs. Their periodograms display broad frequency groups and/or narrow isolated peaks, often superimposed over red noise. The detected signals appear at low frequencies, with few cases of significant signals beyond 5\,d$^{-1}$ (and all of them are faint). The signal amplitudes, and sometimes the frequency content, change with time, even in the absence of outburst events. On the basis of their optical photometric variability, \gc\ stars reveal no distinctive behaviour and thus appear similar to Be stars in general. 
\end{abstract}

\begin{keywords}
stars: early-type -- stars: emission line,Be -- stars: massive -- stars: variable: general 
\end{keywords}

\section{Introduction}
Massive stars may display various spectral peculiarities, which is expressed by specific letters being appended to their spectral types. The Be category was introduced to gather stars presenting Balmer hydrogen lines in emission. This emission is now understood as arising in a circumstellar decretion disk \citep[for a review, see][]{riv13}. \gc\ has long been considered as the prototype of this category, but its high-energy properties (high effective temperature and luminosity intermediate between X-ray binaries and ``normal'' OB stars) appeared at odds with usual observations of massive stars \citep{jer76,mas76,smi16}. With time, more than twenty other Oe/Be stars were found to display these peculiarities \citep{naz20x}, leading to the definition of a ``\gc '' subclass amongst Be stars (not linked to the old optical photometric ``GCAS'' category used in GCVS). The defining criteria of this subclass \citep{naz18} are a Be spectral type in the optical range and a thermal X-ray emission with $\log(L_{\rm X}(0.5-10.\,{\rm keV})\sim31.6-33.2$ (equivalent to $\log(L_{\rm X}/L_{\rm BOL})$ between --6.2 and --4) and $kT>5$\,keV (equivalent to a ratio between the fluxes in the 2.0--10.0\,keV and 0.5--2.0\,keV bands larger than 1.6 or to $L_{\rm X}(2.0-10.\,{\rm keV})>10^{31}$\,erg\,cm$^{-2}$\,s$^{-1}$) - all X-ray fluxes are after correction for the ISM absorption. When the X-ray data are sufficiently numerous and of high signal-to-noise ratios, two additional characteristics become obvious: the presence of Fe\,K$\alpha$ fluorescence line near the iron complex at 6.7\,keV and variations on both short and long timescales \citep{smi16,naz18}. As a comparison point, O-type stars typically show much softer and fainter thermal X-ray emissions, with $kT\sim 0.6$\,keV and $\log(L_{\rm X}/L_{\rm BOL})\sim -7$ (and B-type stars are even fainter), while X-ray binaries exhibit $\log(L_{\rm X})>34$. As not all Be stars have been observed at X-ray wavelengths, it should be noted that many more cases may remain unidentified at the present time hence the exact incidence of the \gc\ peculiarity remains unknown.

The origin of the \gc\ phenomenon remains debated. Two main scenarios have been proposed, one linked to accretion onto a compact companion \citep{ham16,pos17} and another one linked to magnetic star-disk interaction \citep{smi16}. In this context, it is important to underline that, up to now, the only reported difference between \gc\ analogs and the other Be stars lies in their X-ray emission. However, not many \gc\ analogs have been the subject of thorough (optical, ultraviolet, infrared) investigations hence important evidence has yet to be identified.

In this context, we decided to investigate the optical variability of \gc\ analogs. Indeed, the variability directly relates to the disk or the star itself hence it could pinpoint their specificities, if the \gc\ phenomenon arises from the Be star and/or its disk. In \citet{naz20phot}, we already reported the detection of coherent variations from $\pi$\,Aqr and HD\,110432 (BZ\,Cru) while \citet{bor20} analyzed the case of \gc\ itself. We now extend the study to a larger set of \gc\ analogs. The homogeneous dataset provided by \te\ will constrain the overall characteristics of these stars and help comparing them with other Be stars. Sections 2 and 3 will present the data and their analysis, respectively, while Section 4 puts the results in a broader context. Finally, Section 5 summarizes and concludes this paper.

\section{Data reduction and analysis}

Currently, 25 \gc\ stars are known \citep[and references therein]{naz20x}. We have checked whether they had been observed by the Transiting Exoplanet Survey Satellite (\te) and this was the case of sixteen of them. Amongst these was HD\,110432, whose \te\ lightcurve was already reported in \citet{naz20phot}; this paper focuses on the 15 remaining stars but shows HD\,110432 again for comparison. It may be recalled at this point that \te\ has a bandpass centered on the classical $I_c$ filter, but broader as it spans 6000--10\,000\,\AA. It thus records the red and near-infrared emissions of stars.

Four targets were observed at 2\,min cadence, and the associated lightcurves were directly downloaded from the MAST archives\footnote{https://mast.stsci.edu/portal/Mashup/Clients/Mast/Portal.html}. Only the corrected lightcurves with the best quality (quality flag=0) data were considered.

For the remaining eleven stars, only \te\ full frame images (FFI) with 30\,min cadence are available. Individual lightcurves were extracted for each target using the Python package Lightkurve\footnote{https://docs.lightkurve.org/}. Aperture photometry was done on image cutouts of 50$\times$50 pixels and using a source mask defined by pixels above a given flux threshold (10\,e$^-$\,s$^{-1}$ for most stars, increased to 15\,e$^-$\,s$^{-1}$ for SAO\,49725 and \gc\ and to 20\,e$^-$\,s$^{-1}$ for V2156\,Cyg and HD\,119682, to avoid faint neighbours). \gc\ appears saturated in \te\ images, but the stellar signal is not lost, just spread out on more pixels (the larger aperture may increase the potential contamination from other objects, though). Background, including scattered light, was taken out using a principal component analysis (PCA). Furthermore, the data points with errors larger than the mean of the errors plus their 1$\sigma$ dispersion were discarded. 

The \te\ pixel has a size of 21\arcsec\ and extraction is done over several pixels (the usual aperture has 1--3\,px radius), hence crowding issues may arise. Using Simbad, the 2MASS catalog, and {\it Gaia}-DR2, we examined whether our 15 targets had neighbours of similar brightness within 1\arcmin\ radius. Most stars have neighbours which are 3 to 10 magnitudes fainter, hence their contribution can be considered negligible. Only two stars appear problematic. HD\,90563 lies at 37\arcsec\ of HD\,302793 which has a similar brightness and a similar spectral type: it is thus impossible to determine from which star the light variations come. Belonging to NGC\,5281, HD\,119682 is surrounded by several stars, notably the brighter CPD--62$^{\circ}$3559 (K\,2\,II/III) and HD\,119699 (A\,1\,II). The spectral types of the contaminating objects are in this case different from that of the target, but the \te\ images revealed that the images of these sources are totally blended, with our target clearly not dominating the output. Since they may not be fully representative of the emissions of the \gc\ stars, the lightcurves of these two targets should thus be taken with a great caution.

Whatever the cadence, the raw fluxes were converted into magnitudes using $mag=-2.5\times \log(flux)+14$ (constant is arbitrary). Figure \ref{lc} shows the final lightcurves for each source, including HD\,110432 \citep{naz20phot}. Two stars (TYC\,3681-695-1 and V767\,Cen) display strong long-term variations. Their presence leads to large signals at low frequencies which fully hide other information on the photometric behaviour, hence they need to be taken out for our study. The overall trends were determined by averaging the photometric data of these two stars within a 1\,d sliding window and then subtracted to get detrended lightcurves. For TYC\,3681-695-1, very rapid changes at $BJD<$2\,458\,791.8 were further discarded. Three stars (\gc, SAO\,49725, and V2156\,Cyg) were observed in two sectors and the two lightcurves were combined after comparing their mean flux levels and shifting the second lightcurve to match the mean level of the first one.

All lightcurves (individual, detrended, combined) were then analyzed using a modified Fourier algorithm adapted to uneven sampling \citep{hmm,zec09}, as there is a small gap in the middle of each sector lightcurve. Figure \ref{ps} shows the results of these searches, with a 1\% significance level derived using the formula of \citet{mah11}\footnote{Formally, there exists no analytical formula to estimate the significance level of a periodogram peak in case of uneven samplings. However, the \te\ lightcurves mostly consist of evenly sampled data, with only a few gaps, and the Mahy et al. formula provides a very good approximation in this case.}. In addition, to assess the evolution of the variability, we derived the periodograms in sliding windows of 5\,d duration shifted by steps of 0.5\,d. Finally, to search for faint signals (especially at high frequencies) and for red noise, we subtracted the strongest signals. This was done by first detrending the data using a 5\,d sliding window (except for TYC\,3681-695-1 and V767\,Cen, for which detrending was already done before) and then taking out the 2--5 sinusoids corresponding to the strongest peaks (Fig. \ref{ps}). 

\section{Results}
\subsection{Types of variability}
While there are no two identical periodograms, some general features can be identified in Fig. \ref{ps}. First, the power mostly lies at low frequencies, usually below 3\,d$^{-1}$. Second, signals either appear as rather isolated peaks or form broad frequency groups. Those groups are genuine features, not just a peak and its close aliases, although faint aliases (due to holes in sampling) can of course also be detected (see also spectral windows in Fig. \ref{sw}). 

To further detail the observed variability, we subdivide the periodogram features into five different types (see below). To summarize our results, Table \ref{tab:var} gathers the properties of all stars, listing the significant signals and their properties.

\begin{figure}
  \begin{center}
\resizebox{7.4cm}{!}{\includegraphics{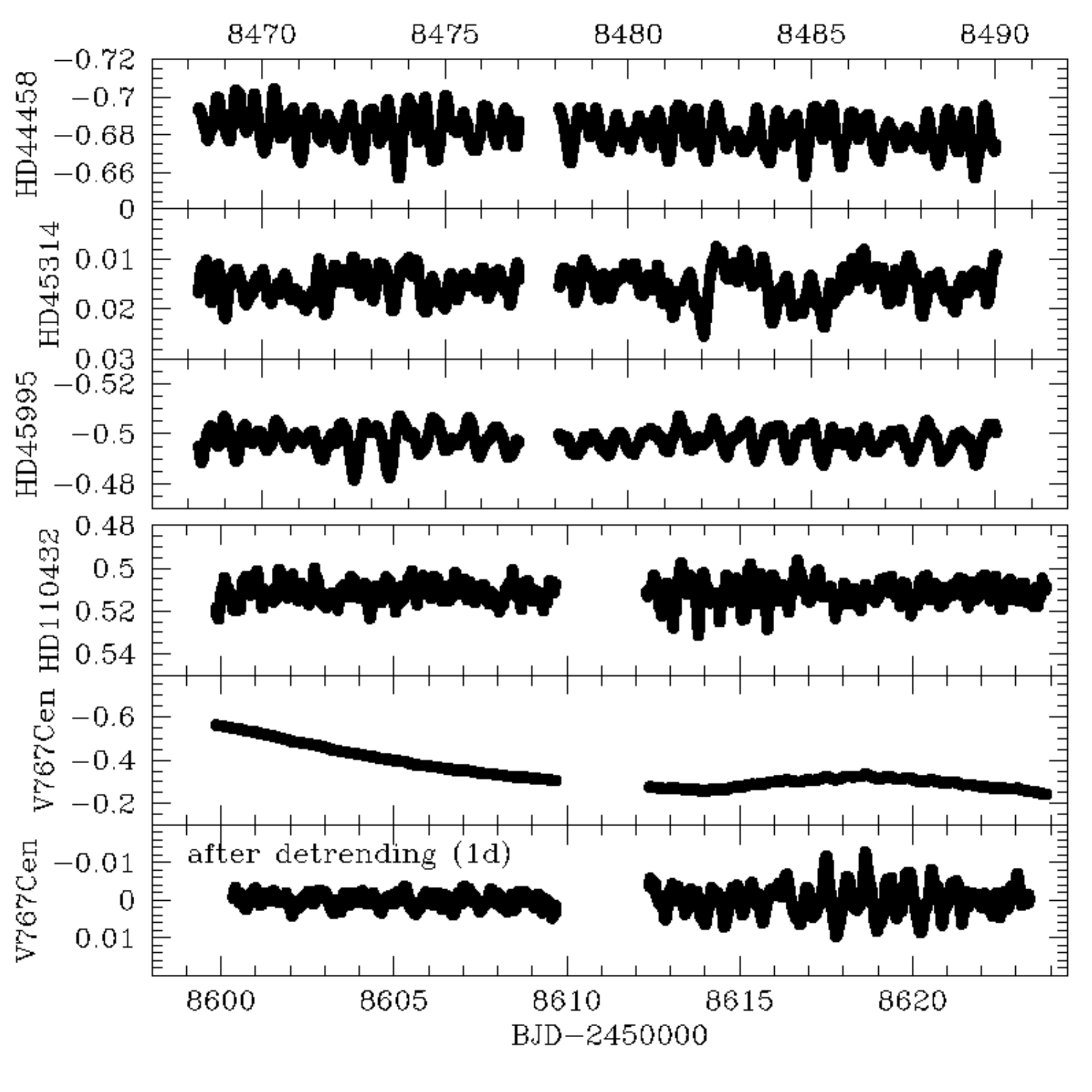}}
\resizebox{7.4cm}{!}{\includegraphics{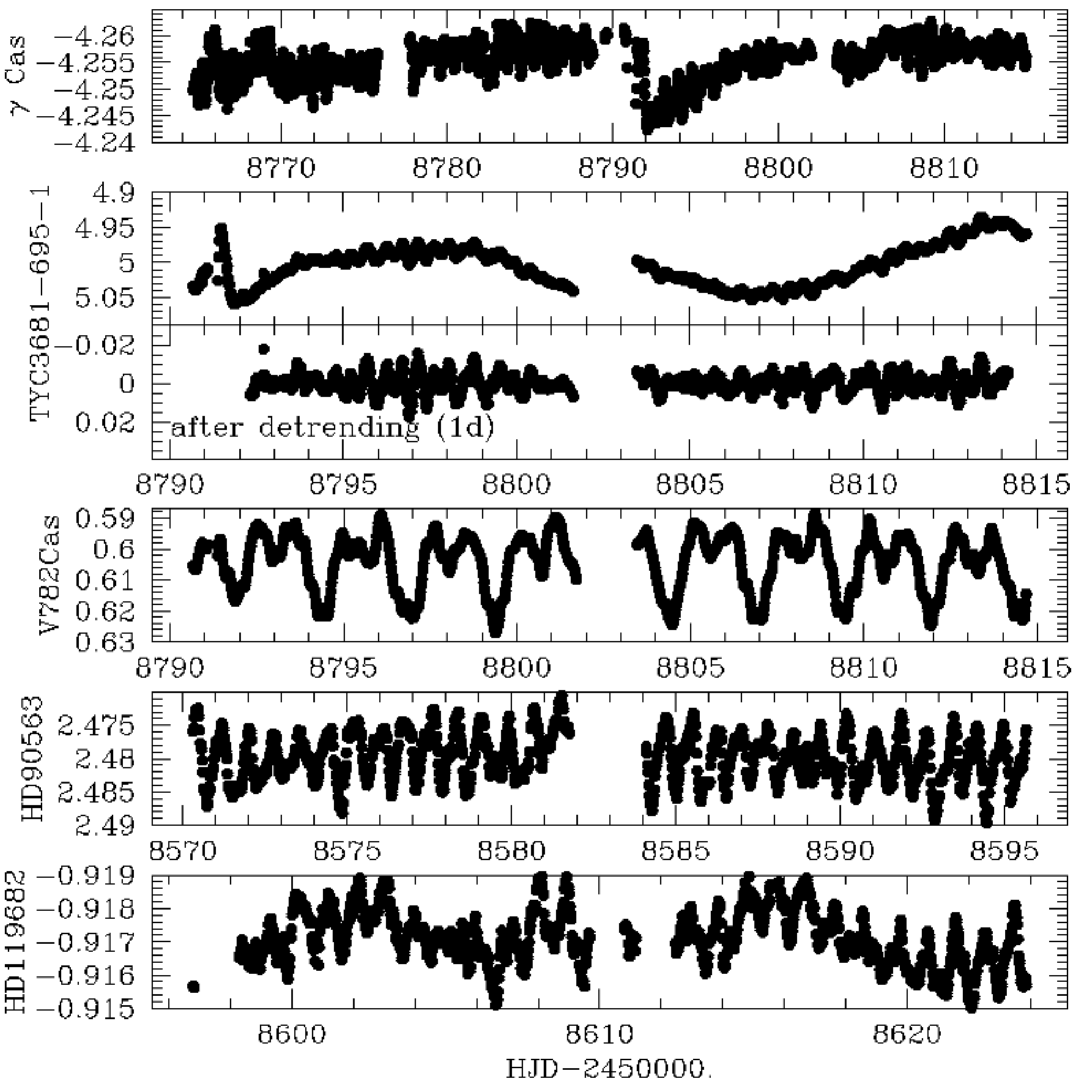}}
\resizebox{7.4cm}{!}{\includegraphics{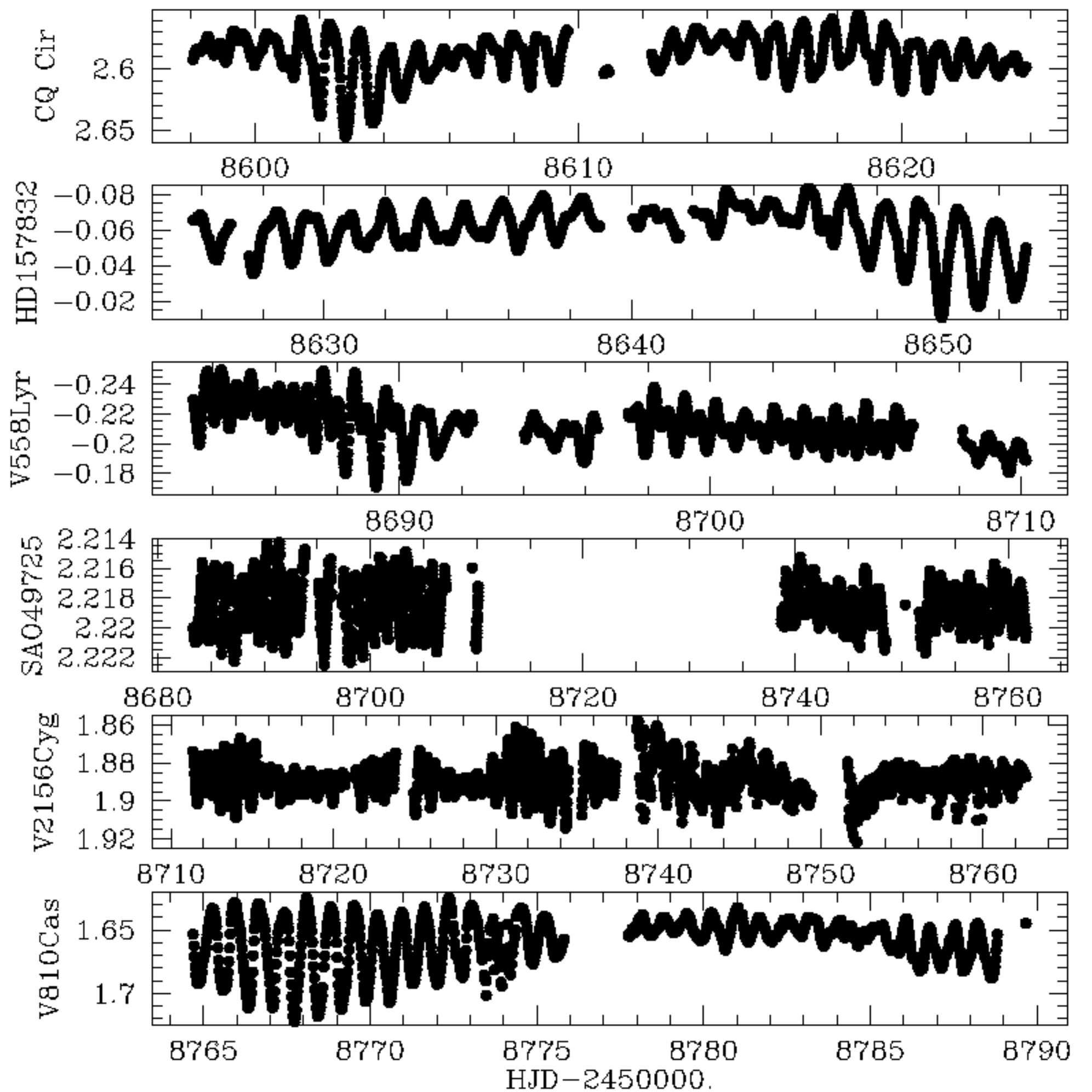}}
  \end{center}
  \caption{\te\ lightcurves of the targets, plus HD\,110432 for completeness; the ordinate provides magnitudes. In the two cases with a large long-term change, the detrended lightcurve is also presented. The top five panels present the 2\,min cadence lightcurves, while the middle and bottom panels show data taken with 30\,min cadence. }
\label{lc}
\end{figure}

\begin{figure}
  \begin{center}
\resizebox{8.5cm}{!}{\includegraphics{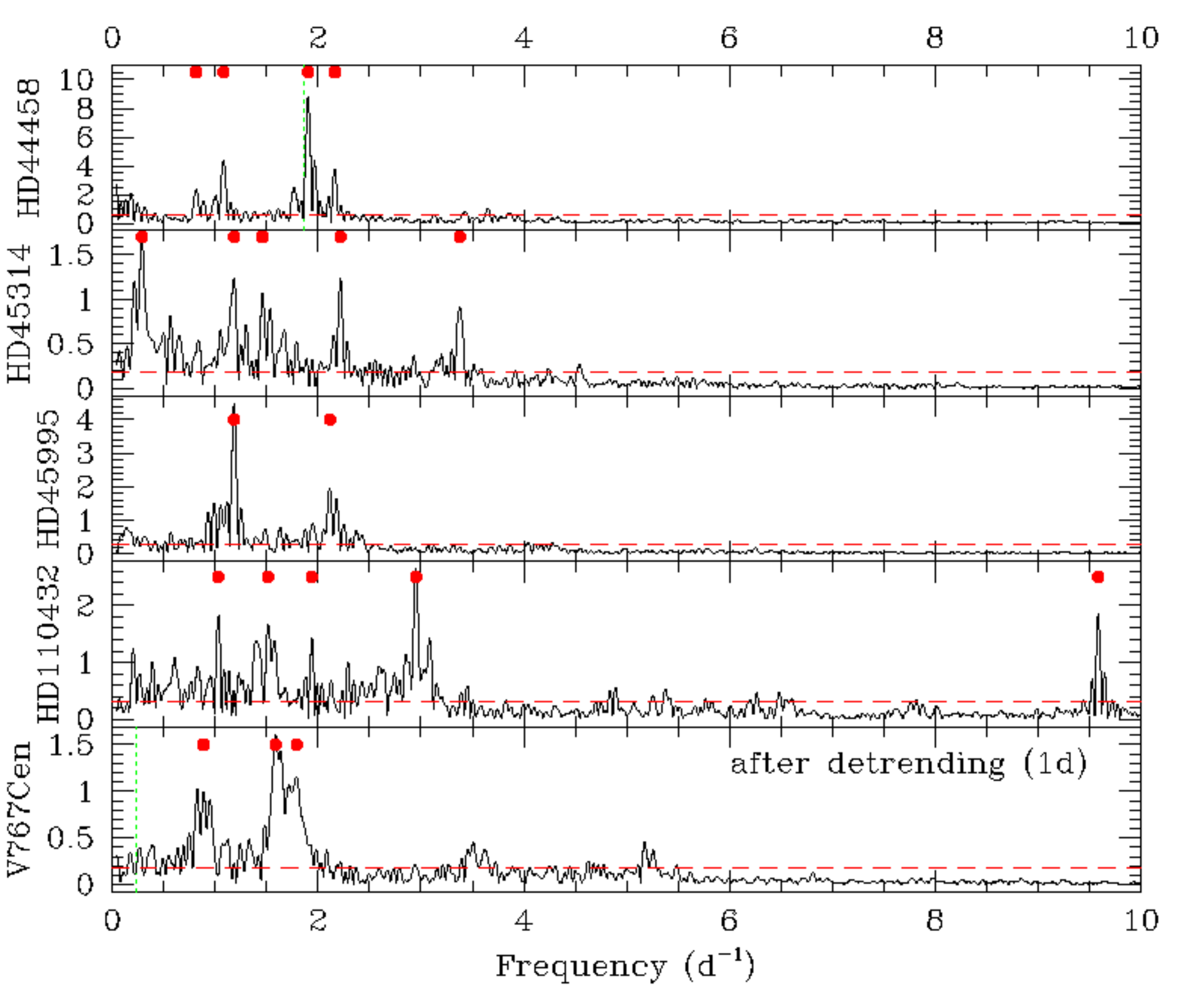}}
\resizebox{8.5cm}{!}{\includegraphics{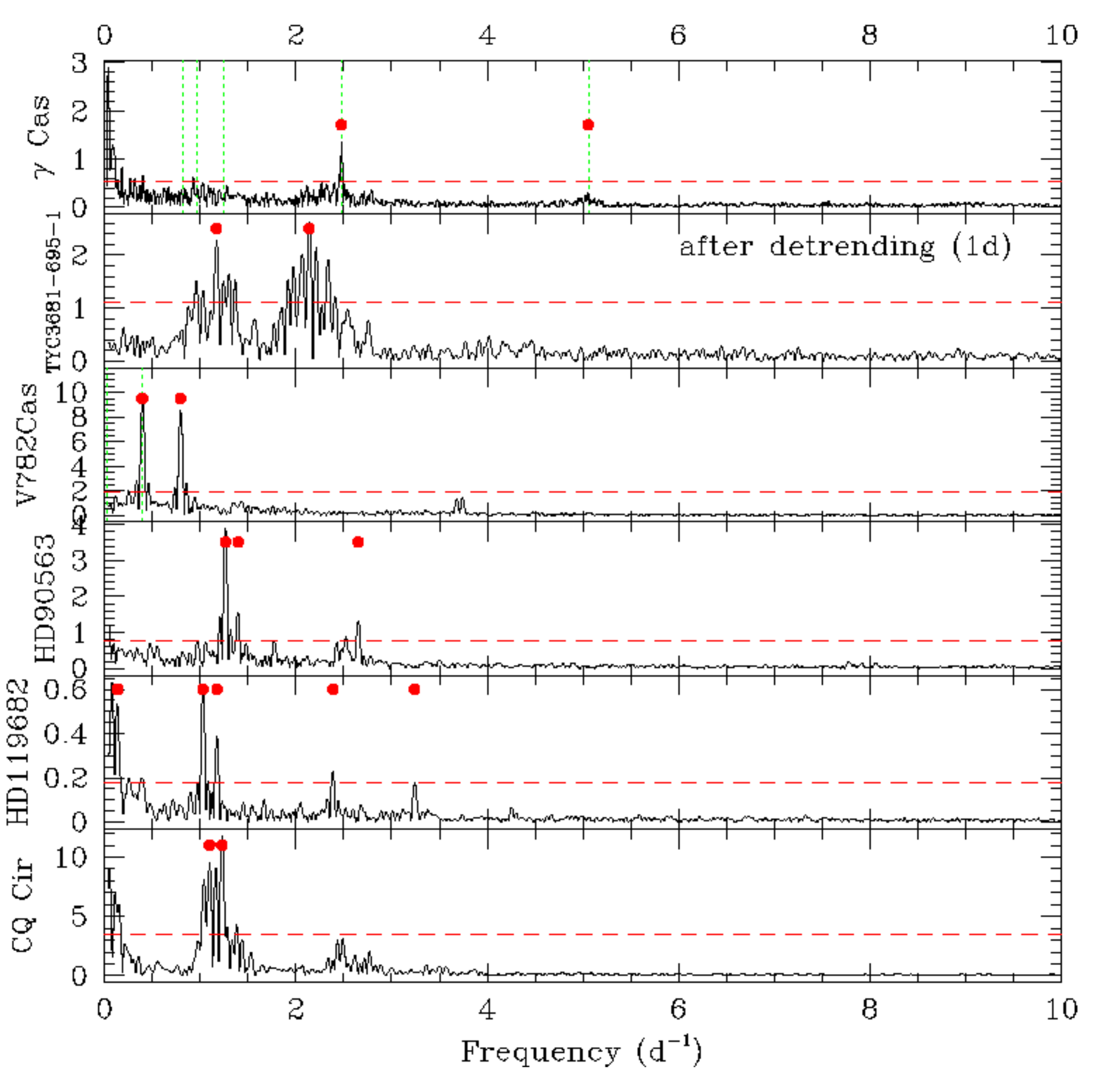}}
\resizebox{8.5cm}{!}{\includegraphics{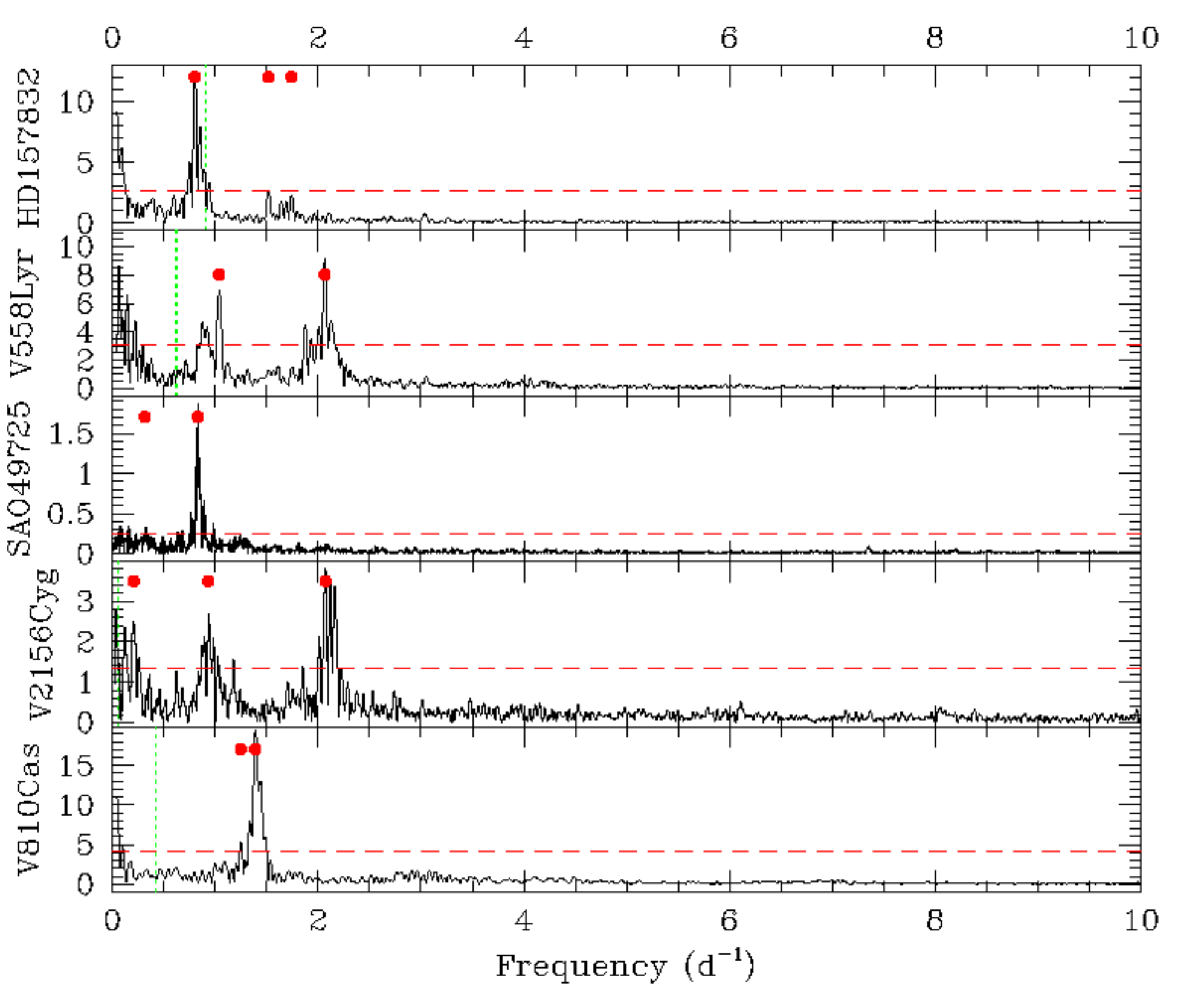}}
  \end{center}
  \caption{Fourier periodograms associated to the lightcurves shown in Fig. \ref{lc}; the ordinate provides sinusoid amplitudes in mmag. For TYC\,3681-695-1 and V767\,Cen, only the periodograms of the detrended lightcurve are presented. The dashed red horizontal line provides the 1\% significance level while dotted green vertical lines show the frequencies of previously reported signals and small red dots indicate the frequencies which were prewhitened to search for specific variability (see Sect. 2). }
\label{ps}
\end{figure}

\begin{landscape}
\begin{table}
\caption{Properties of \gc\ stars, as observed by \te, ordered by R.A.; the list of previously reported frequencies (see Sect. 3.3) is also added for completeness.}    
\label{tab:var}      
\centering       
\small                   
\begin{tabular}{llcrcccccl}        
\hline\hline      
\noalign{\smallskip}
\multicolumn{2}{c}{Star} & Spectral & \multicolumn{1}{c}{$V$} & \te & LTV & FG & \multicolumn{2}{c}{Coherent frequencies (d$^{-1}$)} & Literature $f$\\  
Name & Alt. name & type & \multicolumn{1}{c}{(mag)} & sector(s) & (mmag) & (d$^{-1}$) & ($<5$\,d$^{-1}$) & ($>5$\,d$^{-1}$) & (d$^{-1}$) \\ 
\noalign{\smallskip}
\hline         
\noalign{\smallskip}
\gc\           & HD\,5394  & B0 IVpe        & 2.39 & 17,18 & 10  & 1,2.3,5       & 2.480(D)                         & 5.054,7.572   & 0.822, 0.97, 1.246, 2.479, 5.06\\
TYC\,3681-695-1& ALS\,6507 & B1-2 III/Ve    & 11.36& 18    & 110 & 1.2,2.1       & ---                              & ---           & --- \\
V782\,Cas      & HD\,12882 & B2.5 III:[n]e+ & 7.62 & 18    & $<6$& ---           & 0.396(D),0.796,1.424,3.684,3.744 & ---           & 0.021, 0.398 \\
HD\,44458$^*$  & FR\,CMa   & B1 Vpe         & 5.55 & 6     & 10  & 0.9,1.9       & 0.816(c),1.084,1.904,2.164       & ---           & 1.866 \\
HD\,45314$^*$  & PZ\,Gem   & O9pe           & 6.64 & 6     & $<2$& ---           & 0.288(c),1.184,1.460,2.220,3.380 & 18.080        & --- \\
HD\,45995$^*$  & ALS\,8975 & B2 Vnne        & 6.14 & 6     & $<2$& 1.1,2.1       & 1.184(D)                         & 6.256         & --- \\
HD\,90563      & ALS\,1556 & B2 Ve          & 9.86 & 10    & 3   &2.5,8,10.6     & 1.264(D),1.396(c),1.776,2.656    & ---           & --- \\
HD\,119682     & ALS\,3157 & B0 Ve          & 7.90 & 11    & 1.5 & ---           & 0.140(c),1.032,1.176,2.388,3.244 & ---           & --- \\ 
V767\,Cen$^*$  & HD\,120991& B2 Ve          & 6.10 & 11    & 300 &0.9,1.7,3.5,5.2& ---                              & 6.812         & 0.23474 \\
CQ\,Cir        & HD\,130437& B1 Ve          & 10.04& 11    & 20  &  1.2,2.5      & ---                              & ---           & --- \\ 
HD\,157832     & V750\,Ara & B2 Vne         & 6.66 & 12    & 25  & 0.8,1.7       & 0.804(D),1.520                   & ---           & 0.906 \\
V558\,Lyr      & HD\,183362& B3 Ve          & 6.34 & 14    & 35  & 0.9,2.0       & 1.040,2.068                      & ---           & 0.60972, 0.62625 \\
SAO\,49725     & ALS\,11396& B0.5 III-Ve    & 9.27 & 14,16 & $<3$& ---           & 0.833(D)                         & 7.359,8.194(c)& --- \\
V2156\,Cyg     &           & B1.5 Vnnpe     & 8.91 & 15,16 & 15  & 0.2,0.9,2.1   & ---                              & ---           & 0.048 \\
V810\,Cas      & HD\,220058& B1 npe         & 8.59 & 17    & 15  & ---           & 1.252,1.392(D)                   & ---           & 0.418 \\
\noalign{\smallskip}
\hline   
\end{tabular}

{\scriptsize $^*$ indicates observations taken with 2\,min cadence, (D) dominant frequencies, and (c) peak possibly arising from signal combination. Spectral types are taken from \citet{smi16,naz18,naz20phot} while $V$ magnitudes are from Simbad. LTV corresponds to ``long-term variations'' and their amplitudes were estimated from the trend determined using a sliding average (see Sect. 2 for details). FG stands for ``frequency groups'': because of their broadness, their frequency is approximate and therefore provided with only one decimal. For coherent signals listed in the last two columns, it should be noted that the typical peak widths in periodograms are 0.04\,d$^{-1}$ for observations in one sector or less for multisector cases (0.02\,d$^{-1}$ for \gc\ and V2156\,Cyg, 0.01\,d$^{-1}$ for SAO\,49725); the errors on the peak frequencies will be a fraction of that value (typically one tenth).}
\end{table}
\end{landscape}

\begin{figure}
  \begin{center}
\resizebox{8cm}{!}{\includegraphics{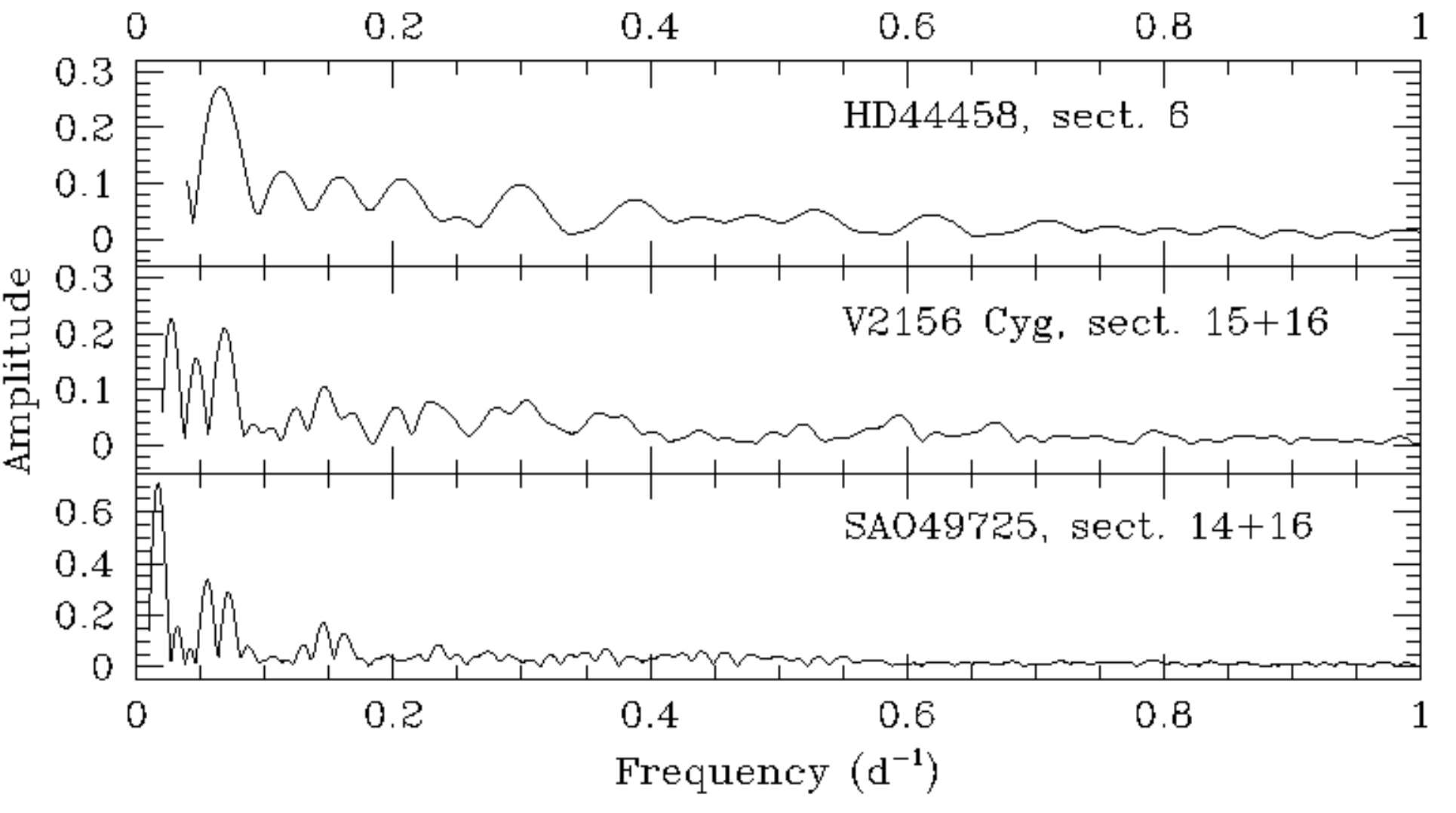}}
  \end{center}
  \caption{Spectral windows derived for one or two sectors, showing expected aliases. }
\label{sw}
\end{figure}

\begin{itemize}
\item {\it Long-term variations:} Be stars are known to vary on long timescales but \te\ photometric data only cover one month per sector, so can only provide a glimpse of the long-term properties of \gc\ stars. In our dataset, large and slow changes are seen in only two stars: TYC\,3681-695-1 and V767\,Cen. The amplitude of these photometric changes are 0.1--0.3\,mag, comparable to what was observed for $\pi$\,Aqr and HD\,110432 in \citet{naz20phot}. Shallower long-term trends ($<0.04$\,mag), resulting in some signal appearing at the lowest frequencies in the periodograms, are seen in nine other stars (see Table \ref{tab:var} for details) but they may be (at least partly) instrumental.
\item {\it Red noise:} A gradual increase in periodogram amplitudes can be seen towards the lowest frequencies, especially in HD\,45314, once the lightcurve has been detrended and cleaned from its main periods. This so-called ``red noise''  actually is stellar in origin, rather than instrumental, as its presence has been reported in several massive stars \citep{blo11,rau19,bow19}. Subsection 3.1.1 provides further details on the properties of the red noise in our targets.
\item{\it Frequency groups:} Rather than isolated peaks, some periodograms show broad groups. In our sample, these frequency groups are dominating the periodograms of TYC\,3681-695-1, V767\,Cen, CQ\,Cir, and V2156\,Cyg. Except for CQ\,Cir, two main groups are present in each star, with the one at a higher frequency having a larger amplitude than the one at a lower frequency. In other stars (especially HD\,45995 and V558\,Lyr, but see details in Table \ref{tab:var}), a strong peak may be accompanied by a dense set of neighbouring fainter peaks, pointing to the presence of a low-amplitude frequency group. 
\item {\it Coherent signals:} When a peak appears strong, isolated, and with a relatively stable frequency (throughout our observations, see Sect. 3.2, and/or comparing with other datasets, see Sect. 3.3), the signal may be assumed to be coherent. Table \ref{tab:var} lists the candidate frequencies. Many stars in our sample appear to possess such signals, with various properties: some peaks are dominant (see below) while others are not; most are at low frequencies while a few others, with low amplitudes, appear at high frequencies (see further details in Sect. 3.1.2); some are in harmonic ratio (see Appendix) while others could be combinations of signals; some appear clearly separated from frequency groups while others appear on top of frequency groups (but clearly dominating them).
  \begin{itemize}
\item {\it Dominant signal:} Some periodograms show a single peak, without any other frequency or with only much fainter other signals. In our sample, this is particularly the case of SAO\,49725 and V810\,Cas, whose periodograms present a single peak. The periodograms of five other stars are dominated by one frequency (with its strong harmonics for V782\,Cas), but fainter signals at different frequencies are also present. In some cases (especially HD\,45995, HD\,157832, SAO\,49725, and V810\,Cas), prewhitening the lightcurves by these dominant frequencies still leaves significant residuals. This could be an effect of the evolving variability (see Sect. 3.3) or reflect the complexity of the underlying signal (actually not a single, pure frequency). 
  \end{itemize}
\end{itemize}

\begin{figure}
  \begin{center}
\resizebox{8cm}{!}{\includegraphics{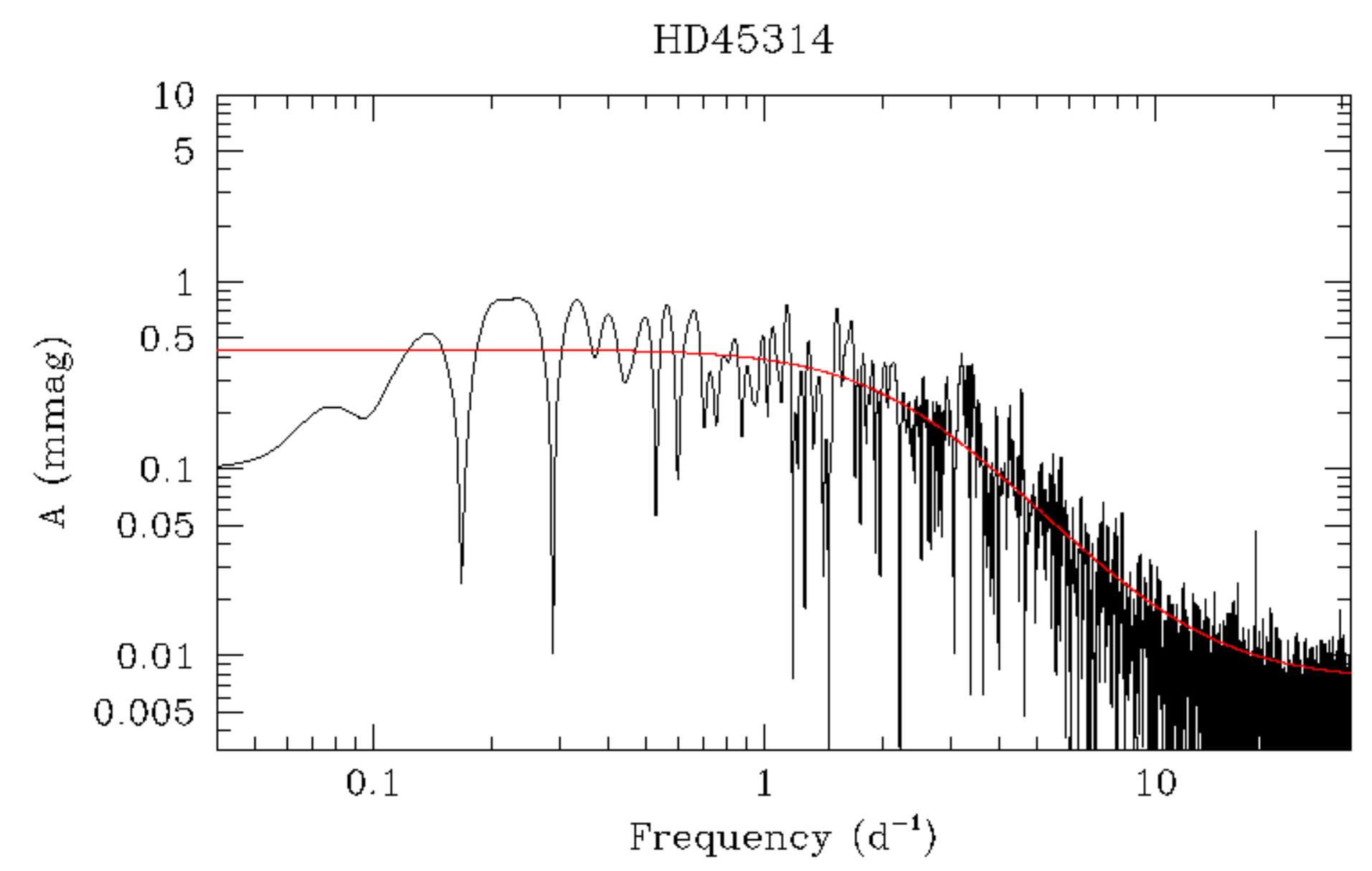}}
\resizebox{8cm}{!}{\includegraphics{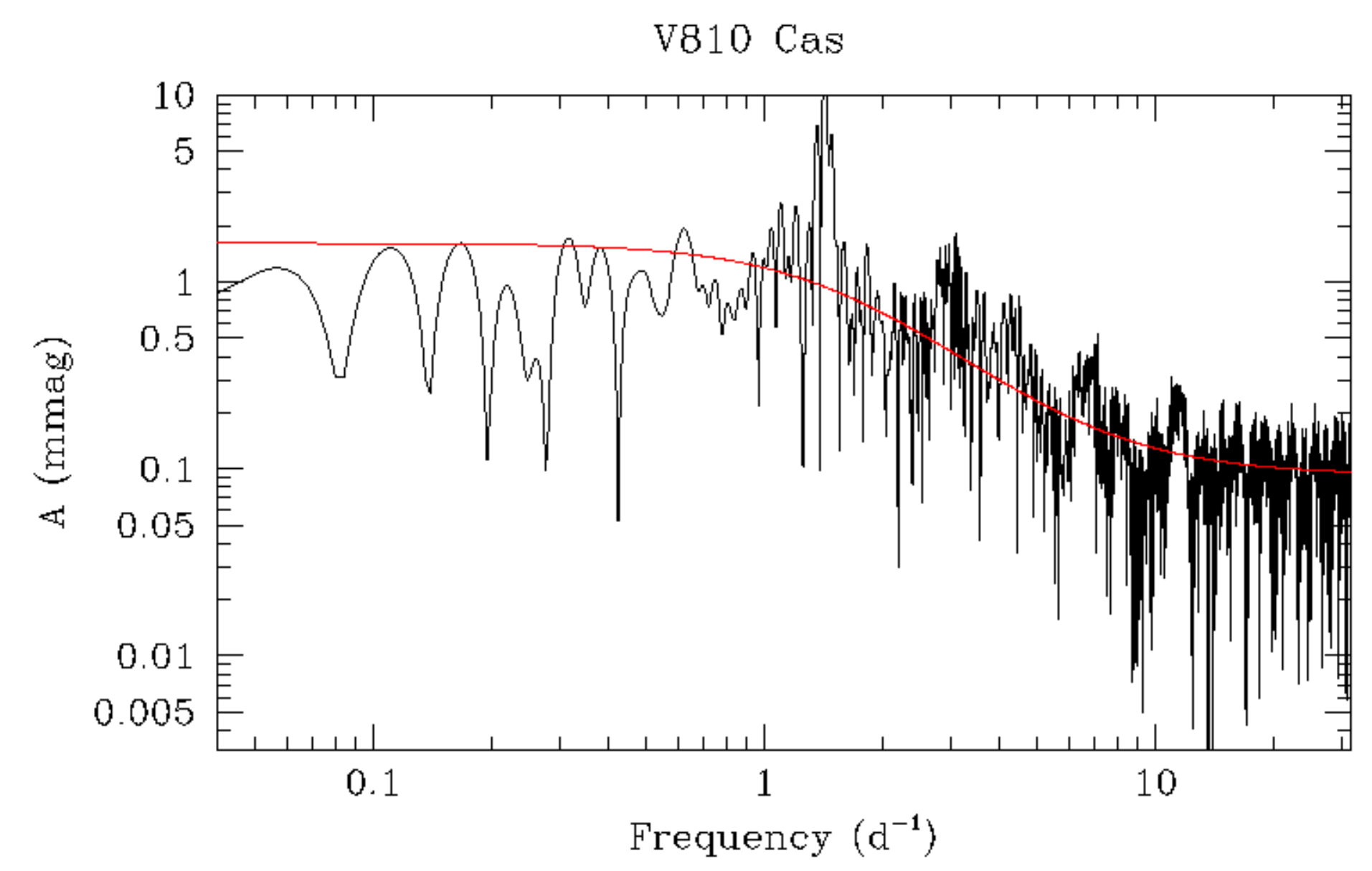}}
 \end{center}
  \caption{Log-log plots of the periodograms of the detrended and prewhitened \te\ photometry of HD\,45314 and V810\,Cas. The red curves correspond to the red-noise relations described by parameters from Table\,\ref{rntable}. }
\label{rnfit}
\end{figure}

\subsubsection{Red noise}
To assess the presence of red noise in the sample of \gc\ stars, we adopt the formalism of \citet{Stanishev} to fit the periodogram with an expression:
\begin{equation}
  A(\nu) = \frac{A_0}{1 + (2\,\pi\,\tau\,\nu)^{\gamma}} + C
  \label{eqrednoise}
\end{equation}
where $A_0$ is the red noise level at nul frequency, $\tau$ the mean lifetime of the structures producing the red noise, $\gamma$ the slope of the linear decrease, and $C$ the white noise level.

\begin{table}
  \caption{White noise and red noise parameters in \te\ data of \gc\ stars, ordered by R.A.. \label{rntable}}
  \begin{tabular}{c c c c c}
    \hline
    Star & $A_0$ & $\gamma$ & $\tau$ & $C$ \\
    & (mmag) &          & (day)  &  (mmag)            \\
    \hline
\gc\           & $0.31 \pm 0.09$ & $1.82 \pm 0.73$ & $0.139 \pm 0.080$ & $0.025$ \\
V782\,Cas      & $0.63 \pm 0.19$ & $1.73 \pm 0.69$ & $0.152 \pm 0.098$ & $0.016$ \\
HD\,44458$^a$  & $0.47 \pm 0.07$ & $2.34 \pm 0.59$ & $0.051 \pm 0.013$ & $0.013$ \\
HD\,45314      & $0.43 \pm 0.04$ & $2.44 \pm 0.42$ & $0.070 \pm 0.010$ & $0.007$ \\    
HD\,90563$^a$  & $0.32 \pm 0.09$ & $1.76 \pm 0.71$ & $0.105 \pm 0.059$ & $0.019$ \\    
HD\,119682     & $0.11 \pm 0.04$ & $1.67 \pm 0.67$ & $0.281 \pm 0.163$ & $0.005$ \\
V558\,Lyr      & $1.41 \pm 0.38$ & $1.99 \pm 0.78$ & $0.100 \pm 0.053$ & $0.032$ \\
V810\,Cas      & $1.52 \pm 0.29$ & $2.03 \pm 0.77$ & $0.099 \pm 0.051$ & $0.093$ \\ 
   \hline
  \end{tabular}
{\scriptsize $^a$ Fit performed over the frequency range 0.1 -- 15\,d$^{-1}$.  }
\end{table}

The parameters of this relation were determined from a fit to the periodogram detrended and prewhitened for the most prominent frequencies (see Fig.\,\ref{rnfit} for examples). For most stars, the fit of the red noise parameters was performed over the frequency range from 0.1\,d$^{-1}$ to 10\,d$^{-1}$, and the level of white noise $C$ was determined from the periodogram mean between 12 and 25\,d$^{-1}$ where red noise is negligible. However, in a few cases, the red noise extended to higher frequencies and we thus performed the fit of the red noise up to 15\,d$^{-1}$, and in one case to 20\,d$^{-1}$. In those cases, the white noise level was determined from the mean between 20 and 25\,d$^{-1}$.

Some remarks must be made on this procedure. First, no fit could be obtained for TYC\,3681-695-1 as significant red noise seems undetectable in the periodogram of this star. Second, the prewhitening was imperfect in some cases, especially when complex variability (frequency groups) was present. In those cases, lots of residual signals remain at low frequencies, which bias the red noise estimates. Therefore, the fitting certainly overestimates the red noise for HD\,45995, CQ\,Cir, and V2156\,Cyg (for which the fit was moreover of poor quality) and it somewhat overestimates it for HD\,110432, V767\,Cen, HD\,157832, and SAO\,49725. For all those stars, the actual red noise parameters remain uncertain. Only the parameters fitted to the remaining cases, for which good and reliable fits could be achieved, are presented in Table\,\ref{rntable}. It is important to note that these values are fully compatible with those observed for massive OB stars \citep{Bow20}.

\begin{figure}
  \begin{center}
\resizebox{8.3cm}{!}{\includegraphics{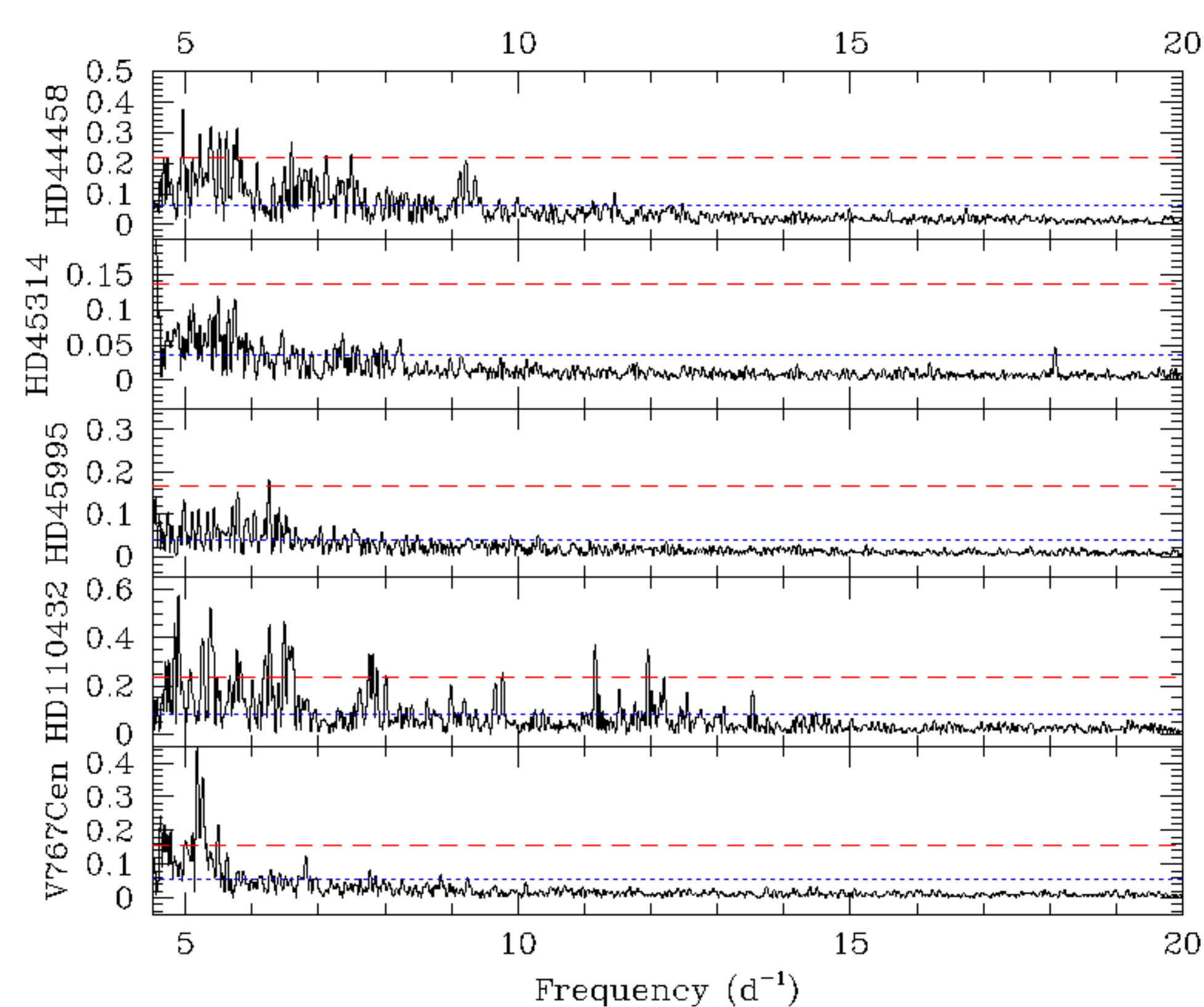}}
\resizebox{8.3cm}{!}{\includegraphics{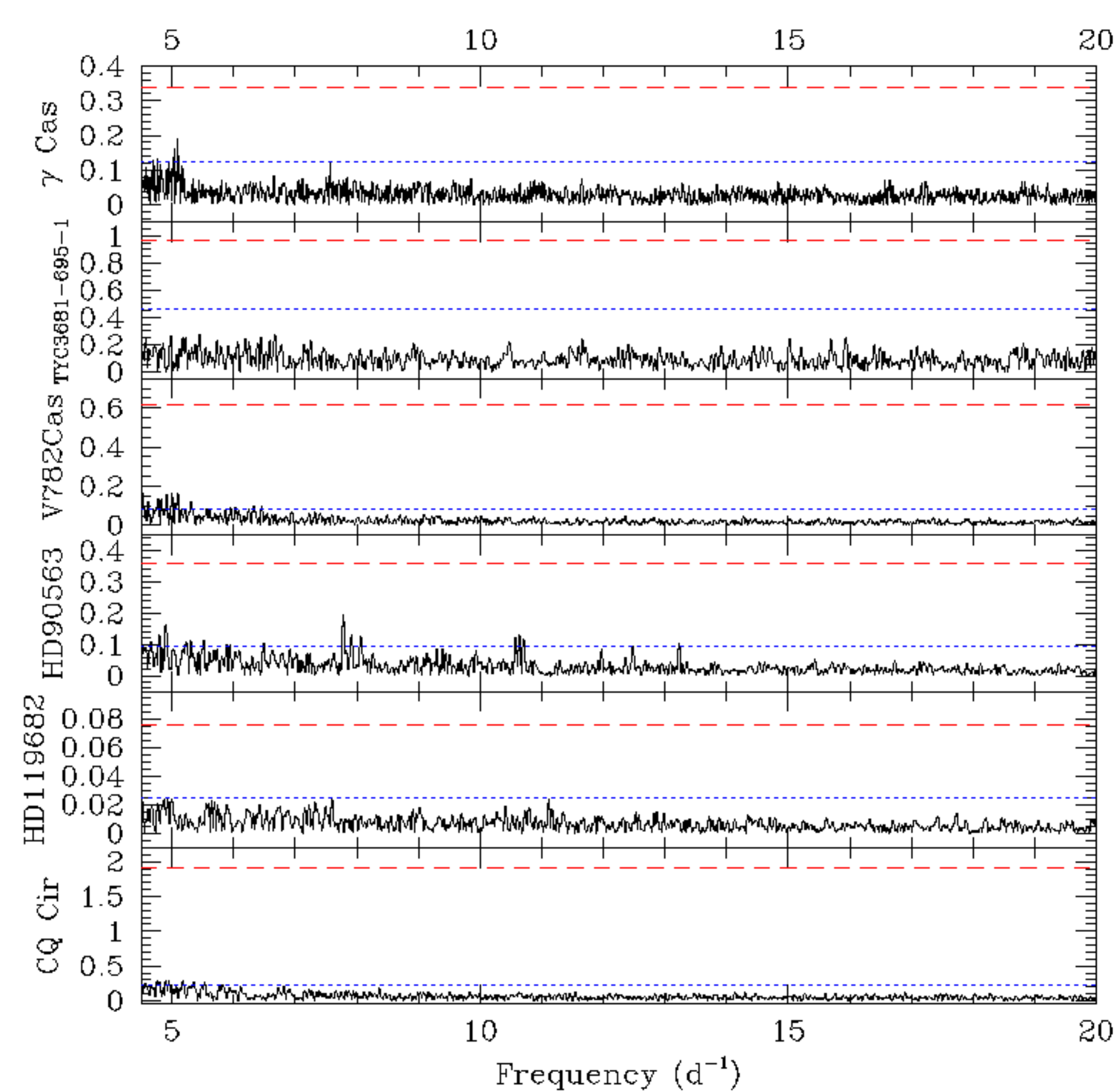}}
\resizebox{8.3cm}{!}{\includegraphics{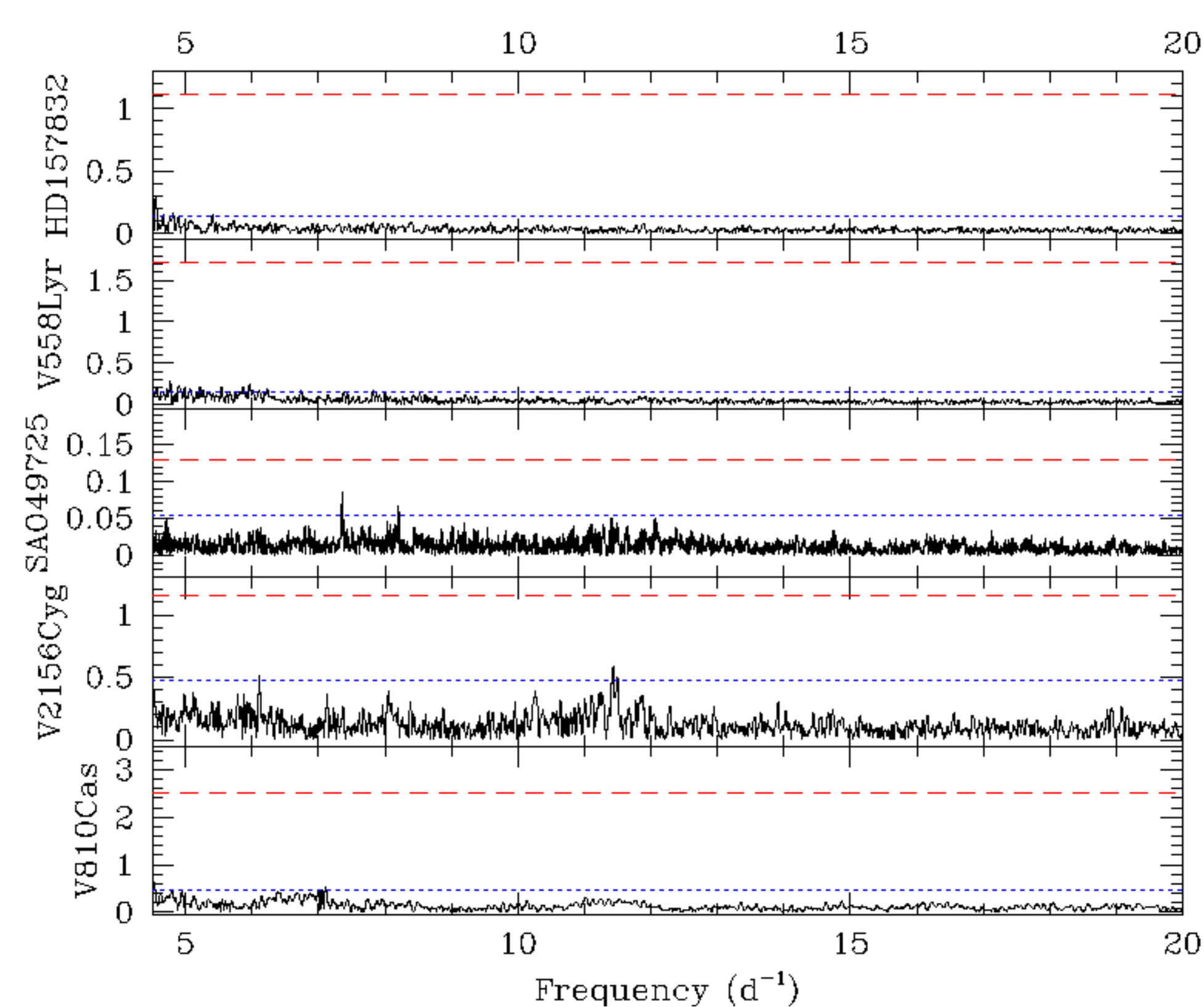}}
  \end{center}
  \caption{Fourier periodograms at high frequencies, after detrending and prewhitening the \te\ lightcurves (see Sec. 2); the ordinate provides sinusoid amplitudes in mmag. In each panel, the dashed red horizontal line provides the 1\% significance level using the formula of \citet{mah11} whereas the dotted blue horizontal line yields 5 times the white noise level (see text for details).}
\label{highf}
\end{figure}

\subsubsection{Presence of high-frequency signals}
All stars studied in this paper display clear variability at low frequencies but are they also varying at high frequencies? In \citet{naz20phot}, the two \gc\ analogs $\pi$\,Aqr and HD\,110432 were found to display coherent high-frequency variations ($f>5\,{\rm d}^{-1}$) with amplitudes of $\sim$1\,mmag. In the {\it SMEI} data of $\pi$\,Aqr, the 2--12\,d$^{-1}$ periodogram was in fact dominated by high-frequency signals (appearing at 7.3, 8.3, 11.6, 11.8\,d$^{-1}$) . In the \te\ data of HD\,110432 (Fig. \ref{ps} and \citealt{naz20phot}), the high-frequency signal at 9.588\,d$^{-1}$ appears as strong as the low-frequency ones, and there are also numerous faint signals at 4.5--7\,d$^{-1}$, near 7.8\,d$^{-1}$ and 11.956\,d$^{-1}$, as well as isolated signals at 11.164\,d$^{-1}$ and 13.528\,d$^{-1}$.

At first sight, nothing similar can be found in the 15 \te\ lightcurves analyzed here (Fig. \ref{ps}). However, low-level signals may be present but be difficult to detect when other signals dominate. The \te\ lightcurves were therefore detrended and prewhitened for the strongest signals to search for them (Sect. 2, Fig. \ref{highf}). Since the 1\% significance level derived using the \citet{mah11} formula considers the full lightcurve scatter, whatever the variability frequency, it underestimates the significance of high-frequency signals when red noise is present or in case of imperfect cleaning at low frequencies (in other words, whenever there remain low-frequency signals). Therefore, we used as an alternative for detection five times the white noise level determined in Sect. 3.1.1 (see Fig. \ref{highf}). 

HD\,44458 presents many low-level signals up to 10\,d$^{-1}$, without clearly isolated peaks (which could be groups or red noise). HD\,45314 displays an isolated peak at 18.080\,d$^{-1}$, while for HD\,45995, a significant and rather isolated peak is present at 6.256\,d$^{-1}$. V767\,Cen displays a significant frequency group near 5.2\,d$^{-1}$, plus an isolated peak at 6.812\,d$^{-1}$. The periodogram of \gc\ shows a peak at 5.054\,d$^{-1}$, which is accompanied by another, even fainter one at 7.572\,d$^{-1}$ (it barely reaches significance). The periodogram of HD\,90563 has small frequency groups near 8 and 10.6\,d$^{-1}$ while SAO\,49725 shows isolated peaks at 7.359 and 8.194\,d$^{-1}$. All these signals reach amplitudes of 0.05--0.4\,mmag, i.e. much less than for HD\,110432 or $\pi$\,Aqr: clearly, our targets do not display strong high-frequency signals but half of them (7 out of 15 if considering groups) display low-level ones.

\begin{figure*}
  \begin{center}
\resizebox{5.8cm}{!}{\includegraphics{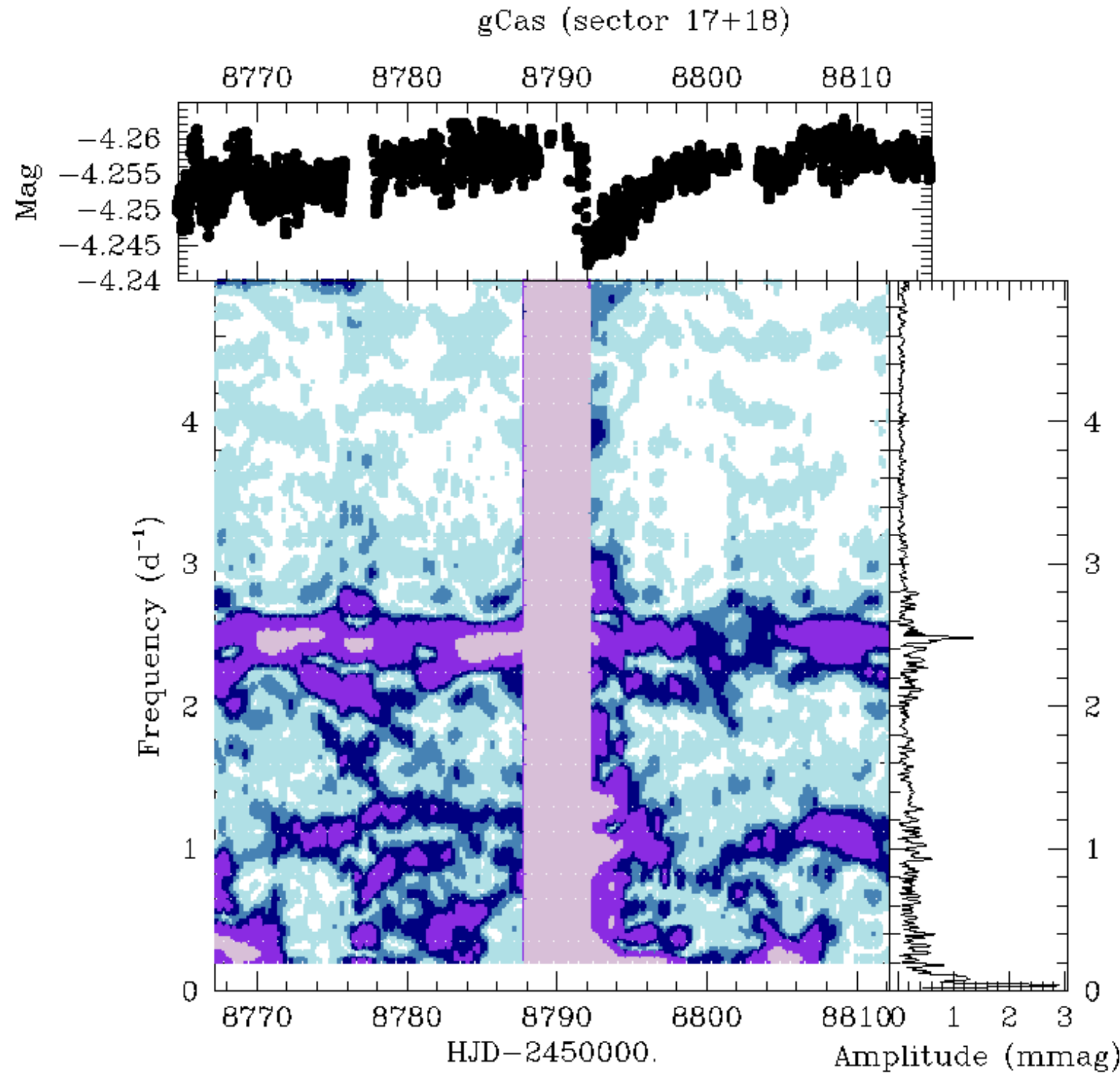}}
\resizebox{5.8cm}{!}{\includegraphics{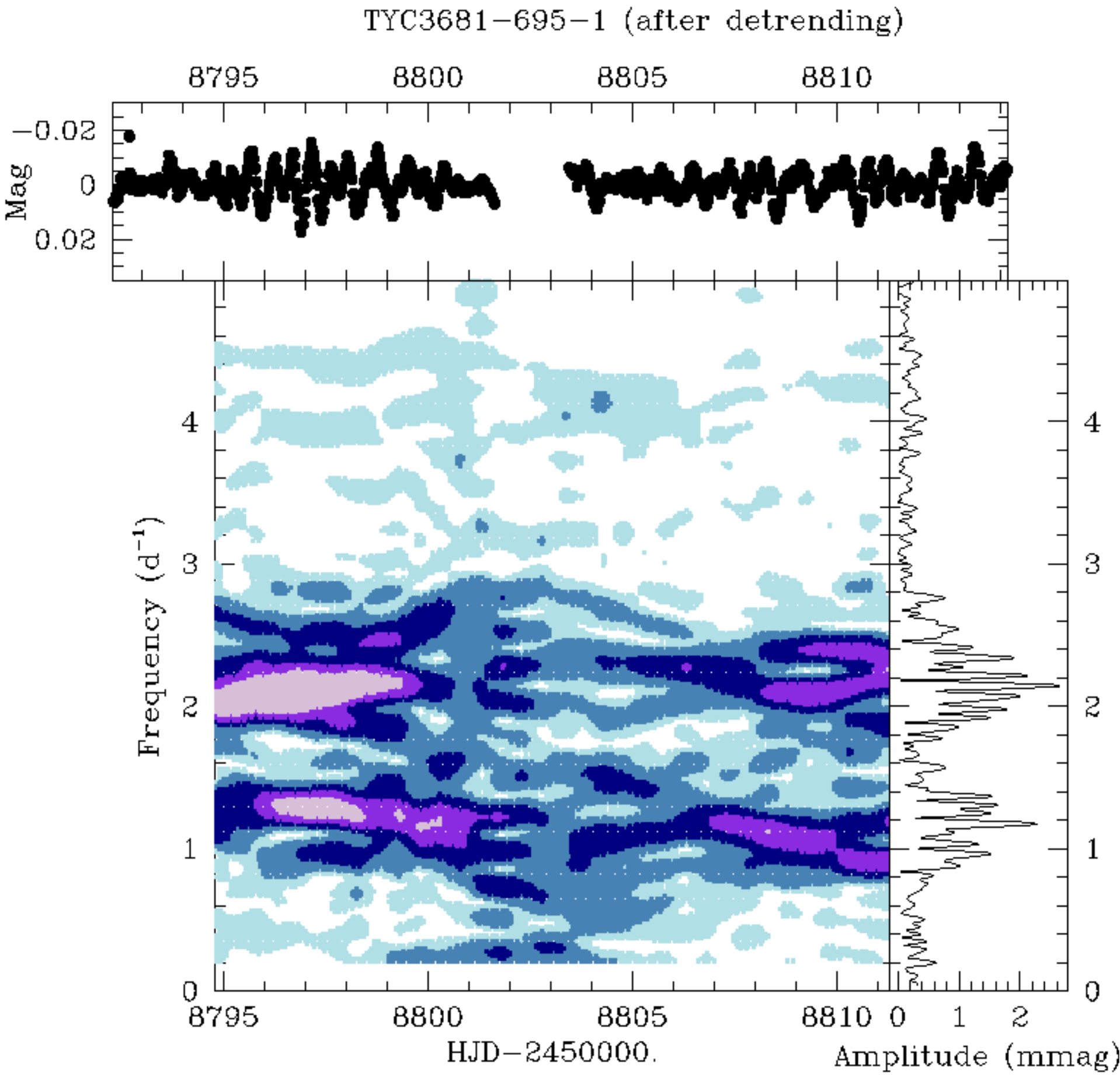}}
\resizebox{5.8cm}{!}{\includegraphics{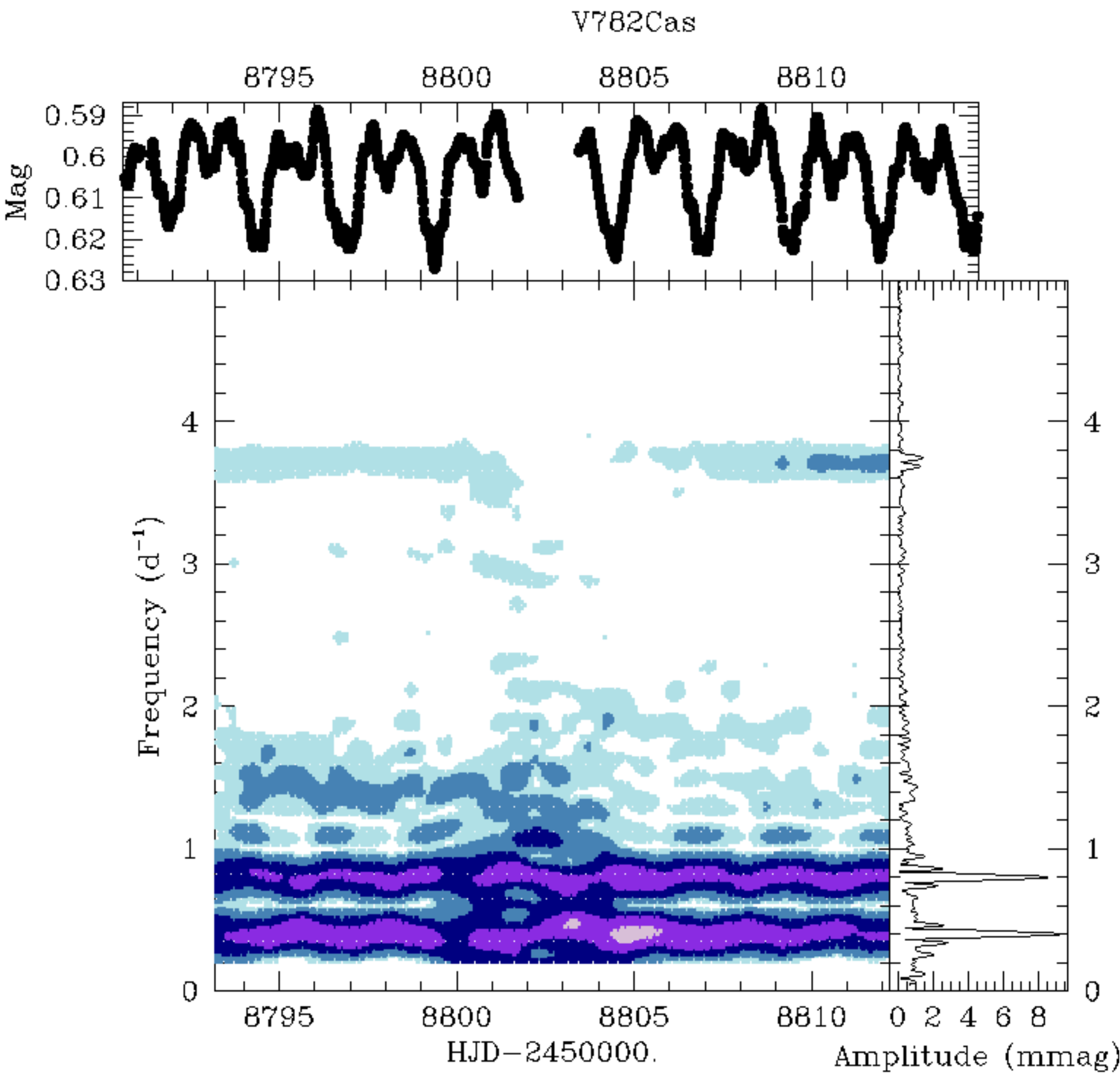}}
\resizebox{5.8cm}{!}{\includegraphics{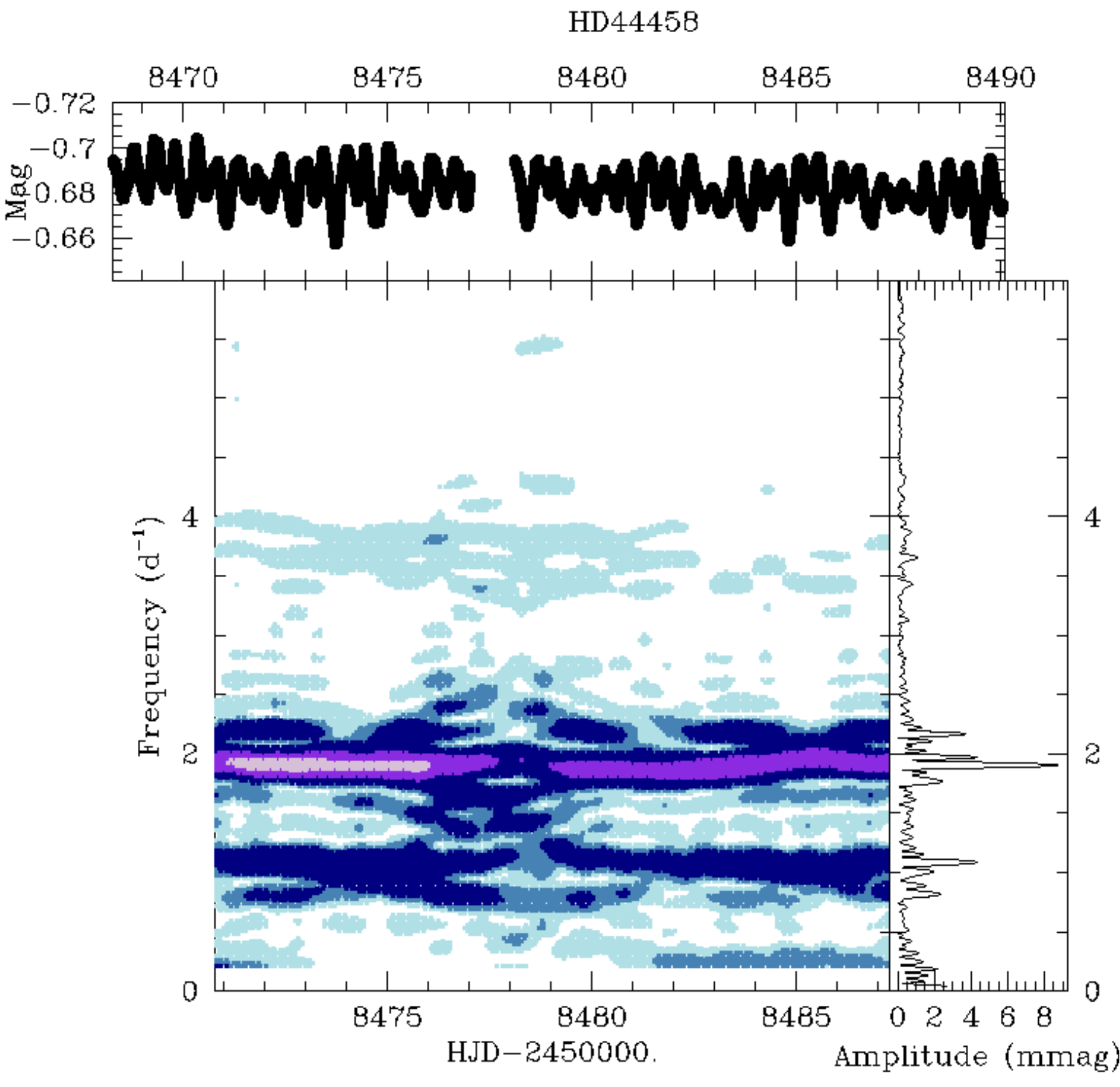}}
\resizebox{5.8cm}{!}{\includegraphics{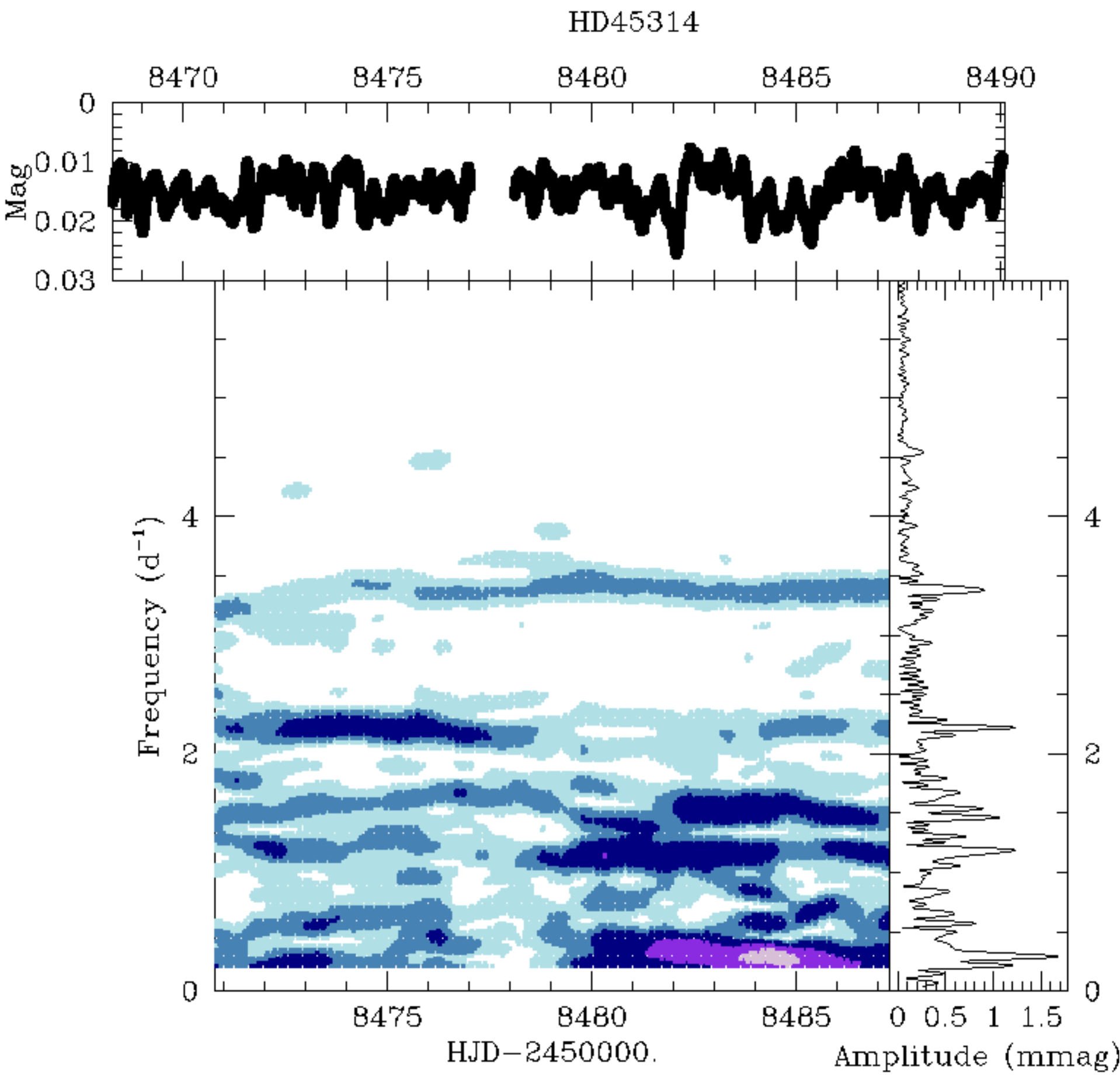}}
\resizebox{5.8cm}{!}{\includegraphics{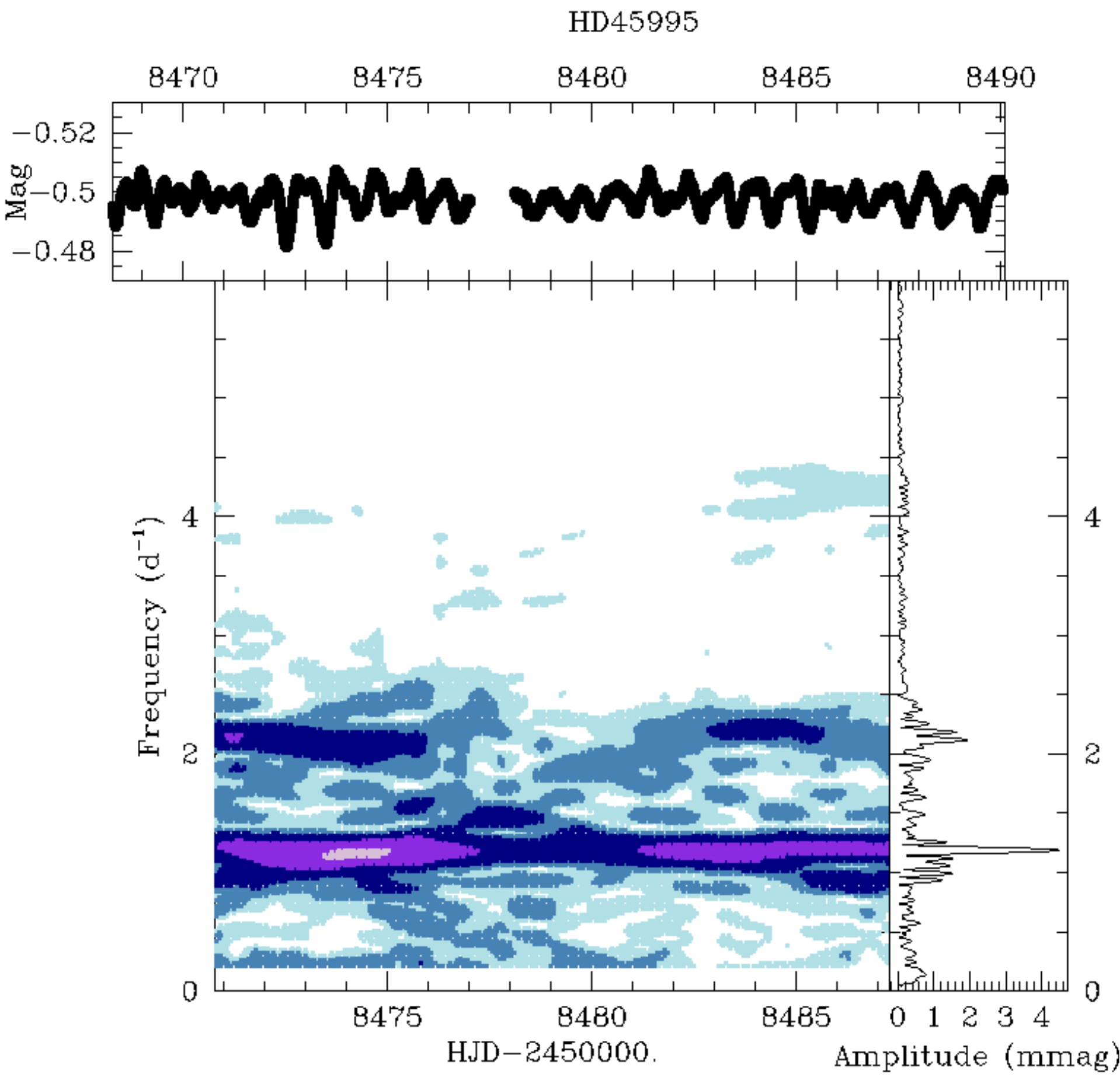}}
\resizebox{5.8cm}{!}{\includegraphics{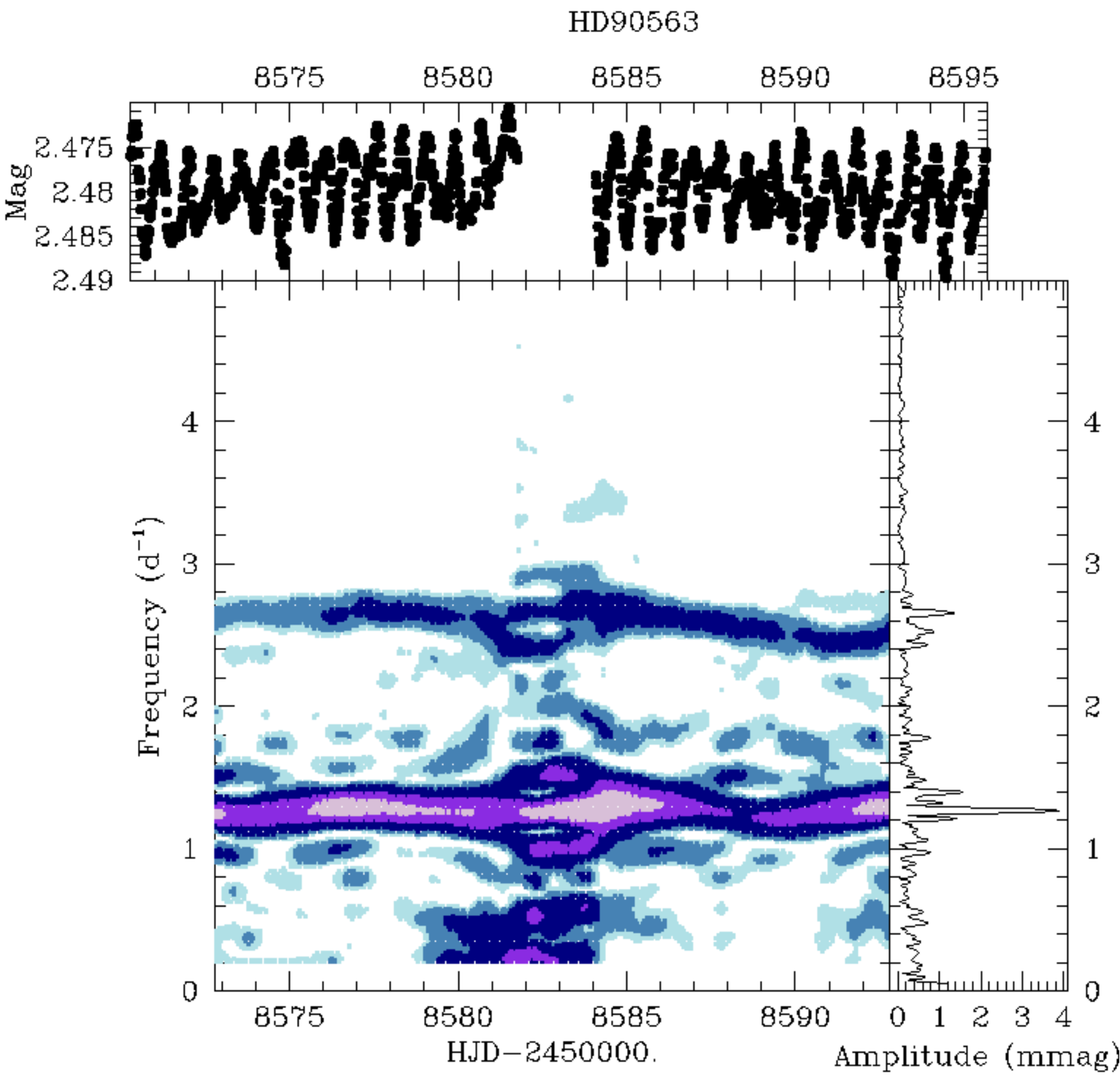}}
\resizebox{5.8cm}{!}{\includegraphics{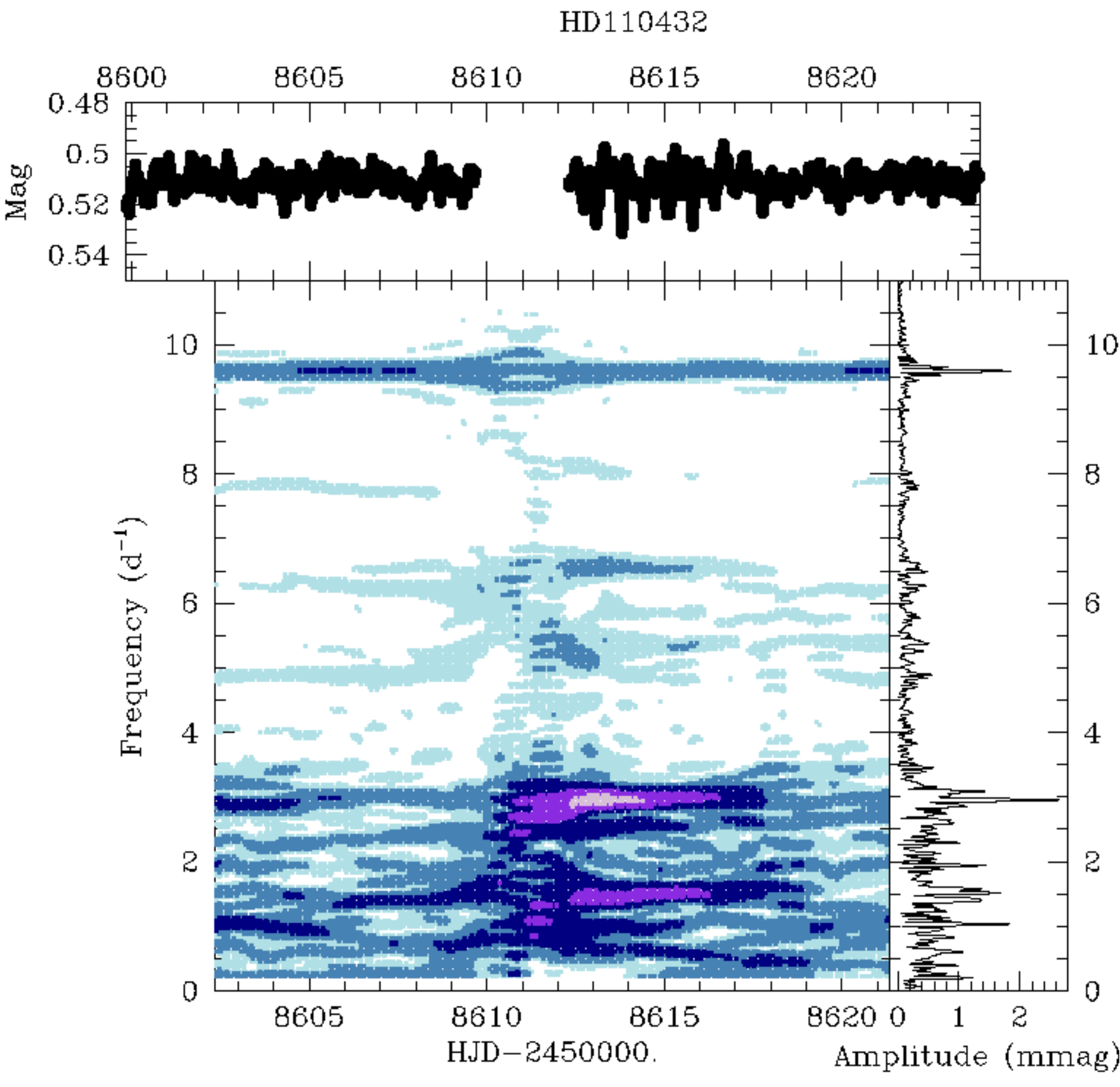}}
\resizebox{5.8cm}{!}{\includegraphics{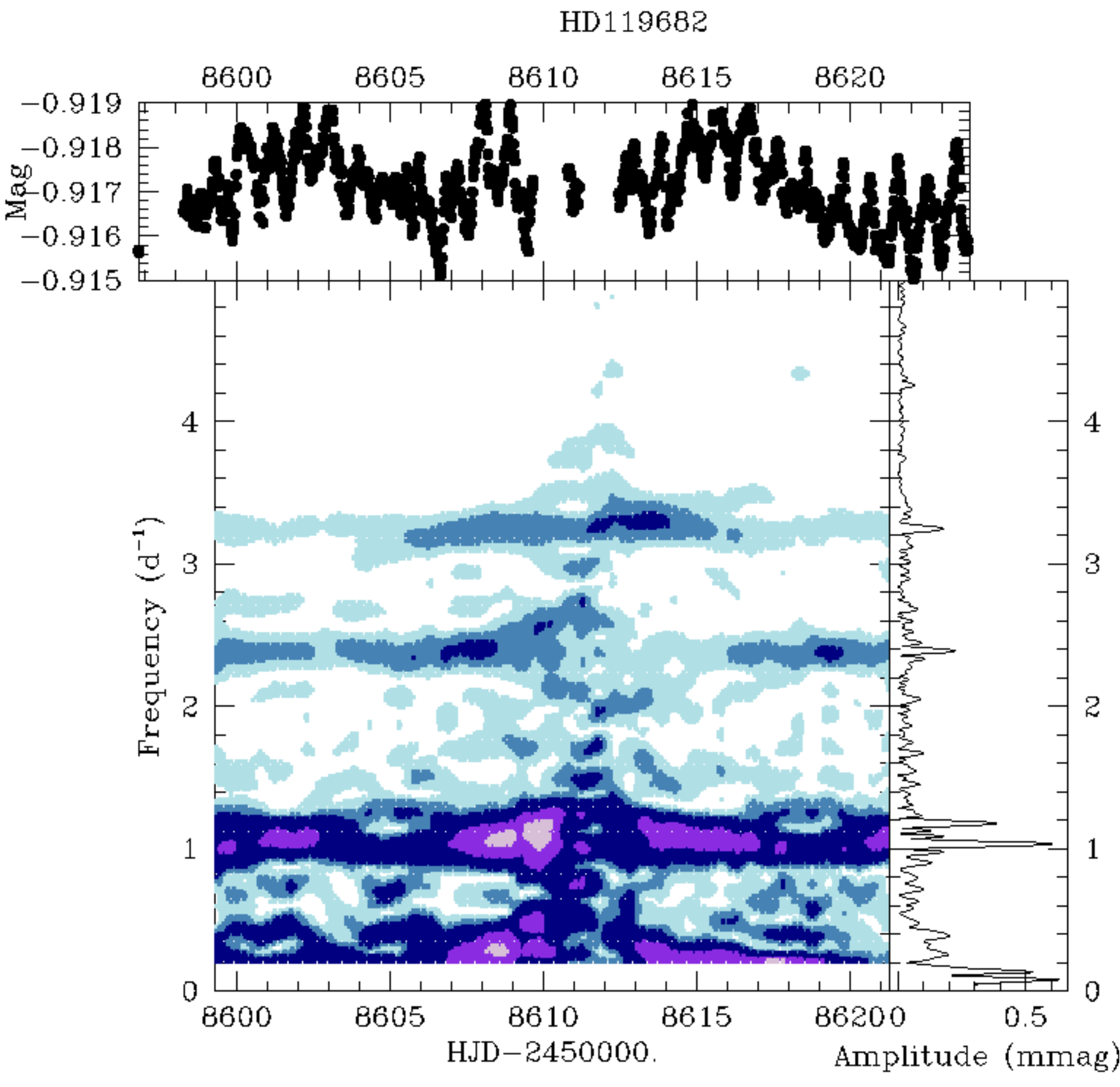}}
\resizebox{5.8cm}{!}{\includegraphics{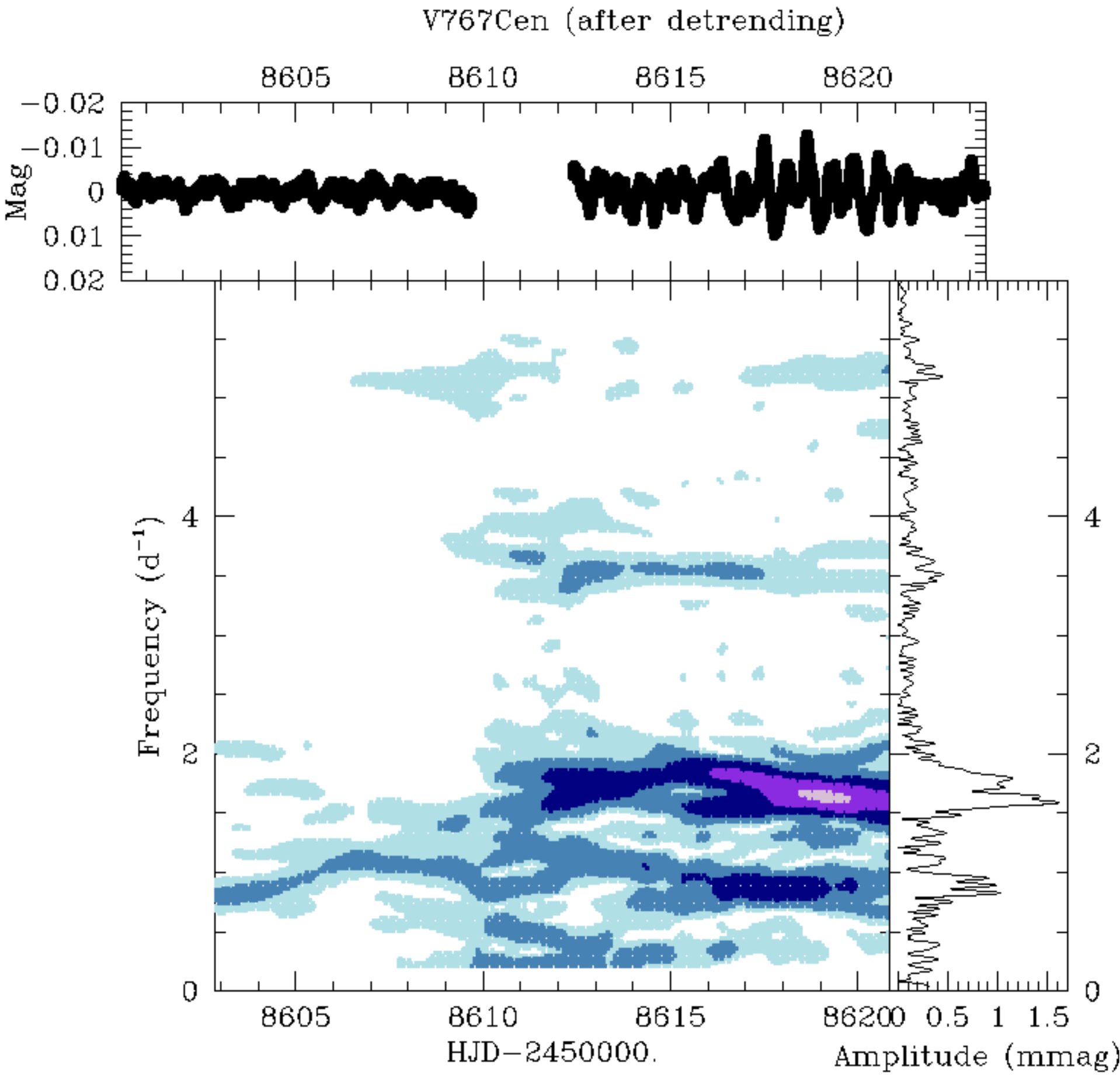}}
\resizebox{5.8cm}{!}{\includegraphics{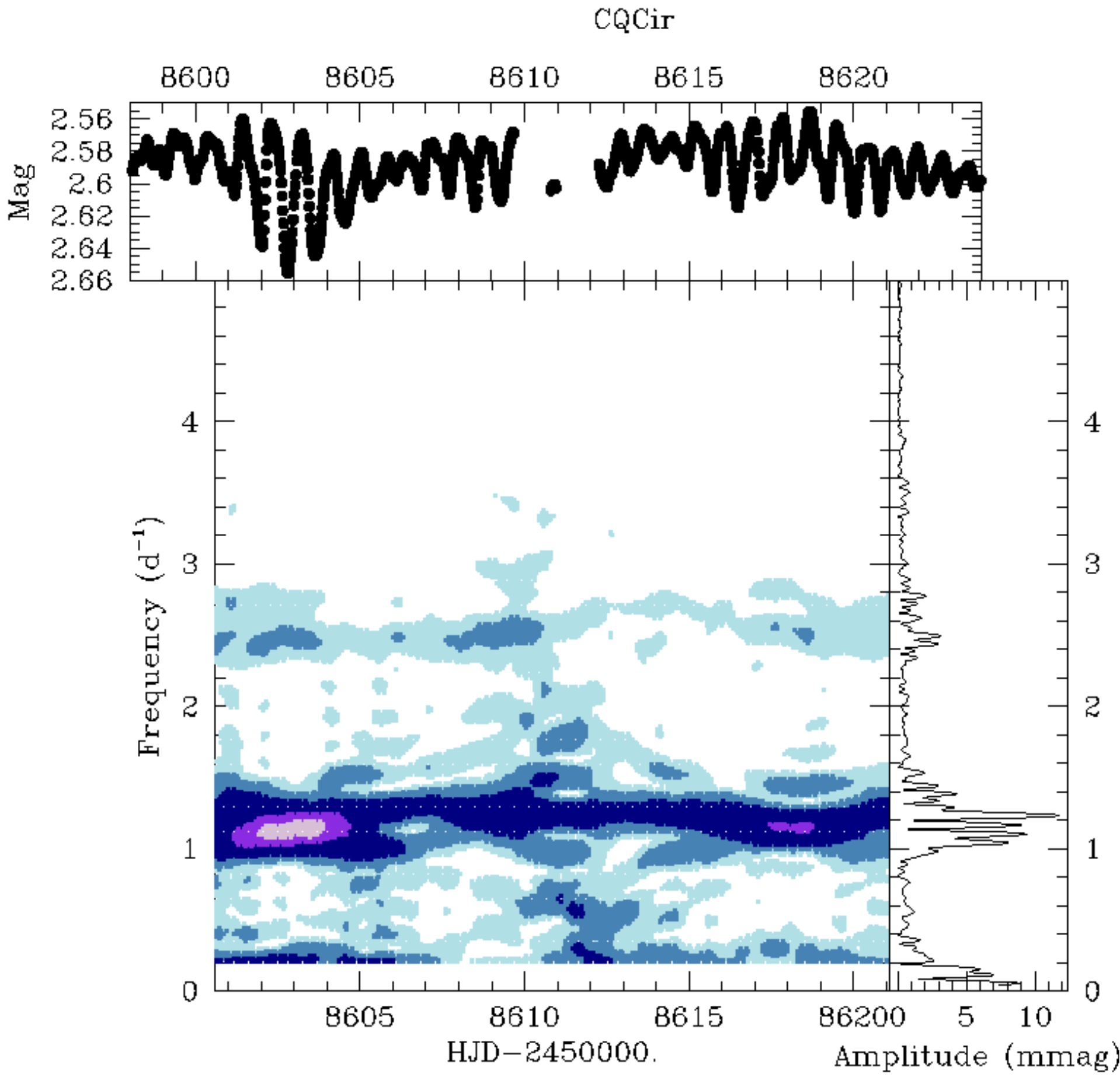}}
\resizebox{5.8cm}{!}{\includegraphics{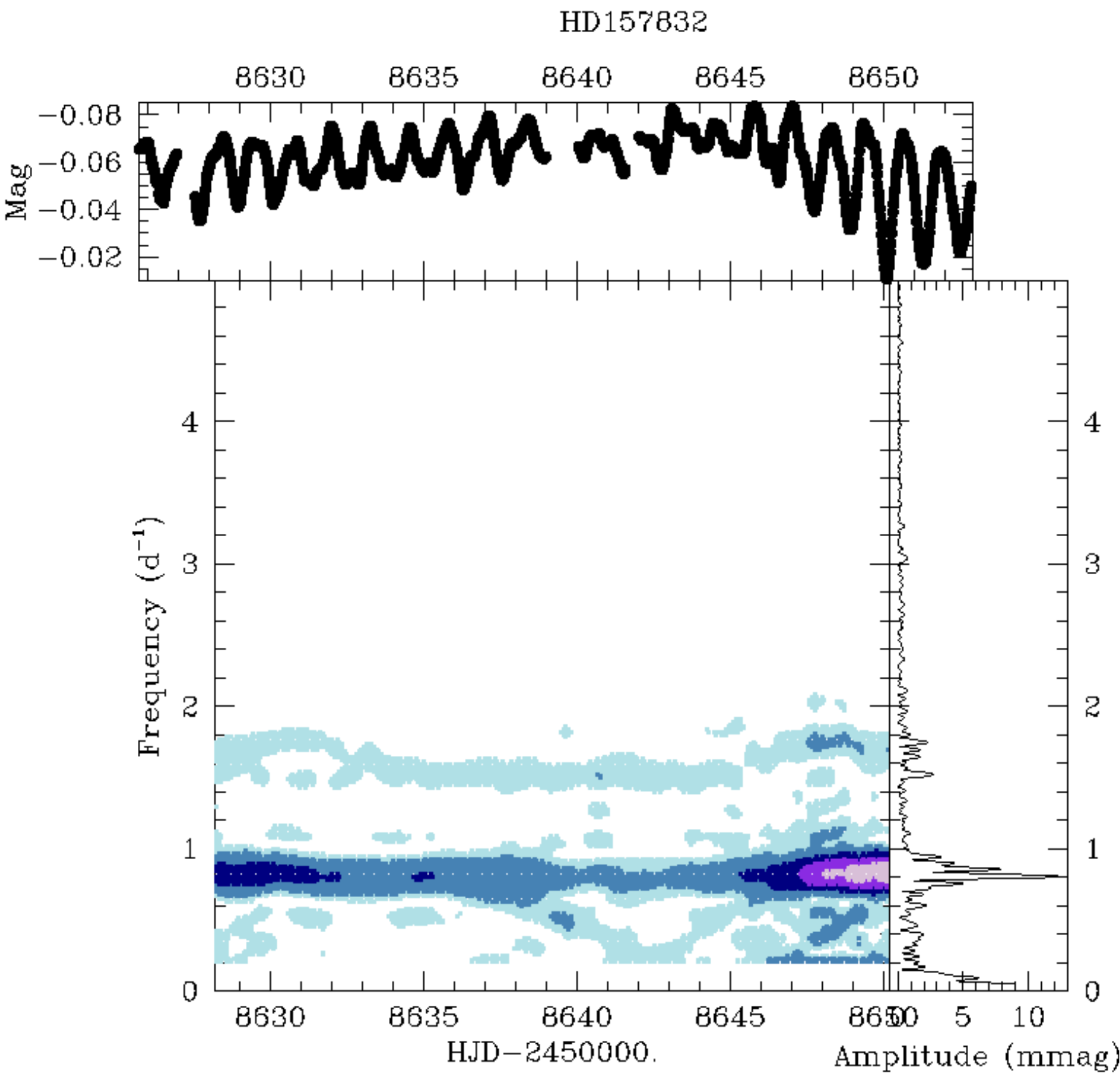}}
  \end{center}
  \caption{Time-frequency diagrams for all targets; the dates in asbcissa correspond to the mid-point of the temporal window used for calculating the periodogram. The lightcurve is displayed on top, whilst the periodogram for the full dataset is shown on the right. For the stars with multi-sector information, the same amplitudes are used for the color levels, to spot changes more easily. For the two stars with strong long-term changes, only the analysis of the detrended lightcurve is provided.}
\label{evol}
\end{figure*}
\setcounter{figure}{5}
\begin{figure*}
  \begin{center}
\resizebox{5.8cm}{!}{\includegraphics{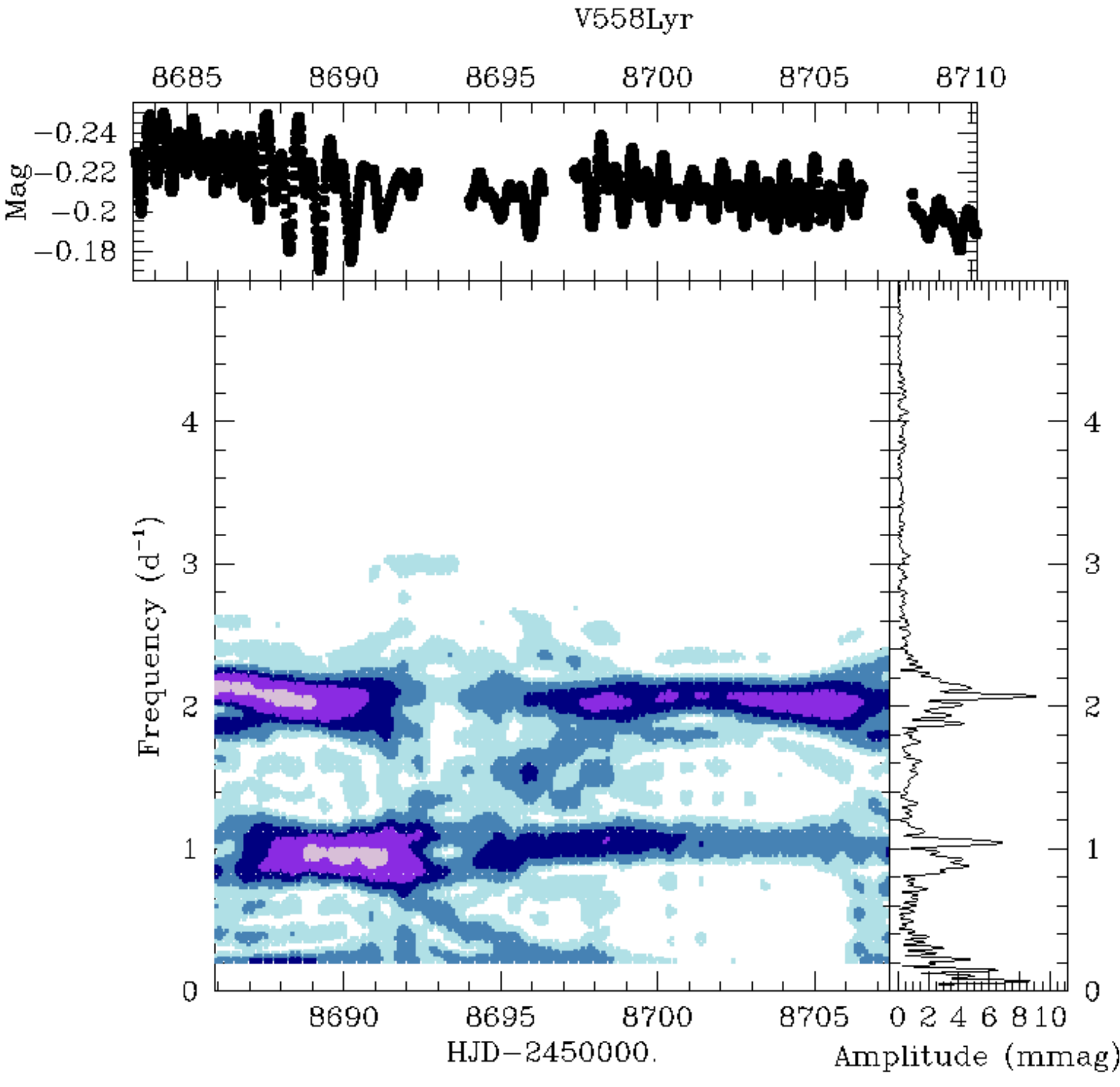}}
\resizebox{5.8cm}{!}{\includegraphics{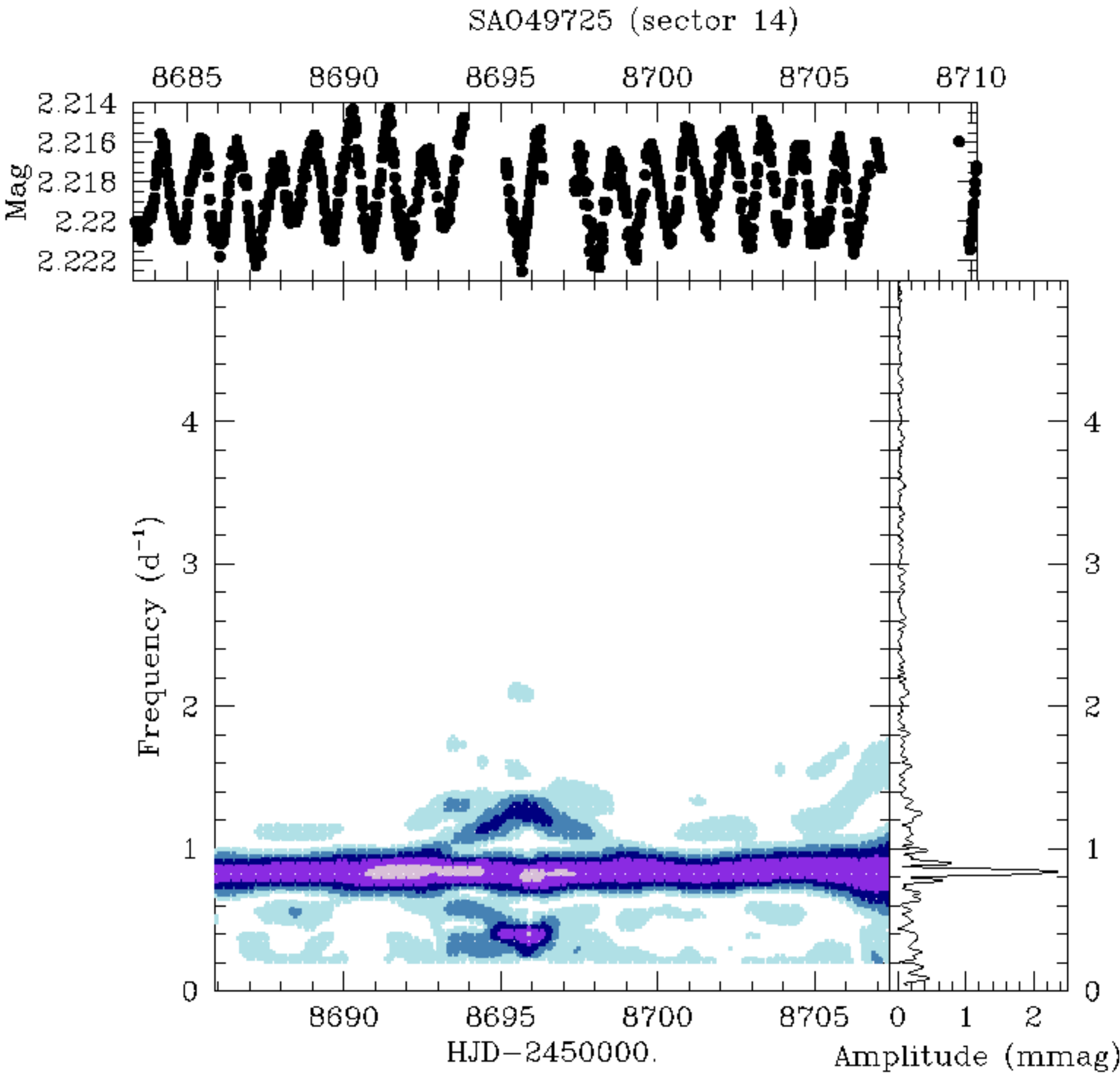}}
\resizebox{5.8cm}{!}{\includegraphics{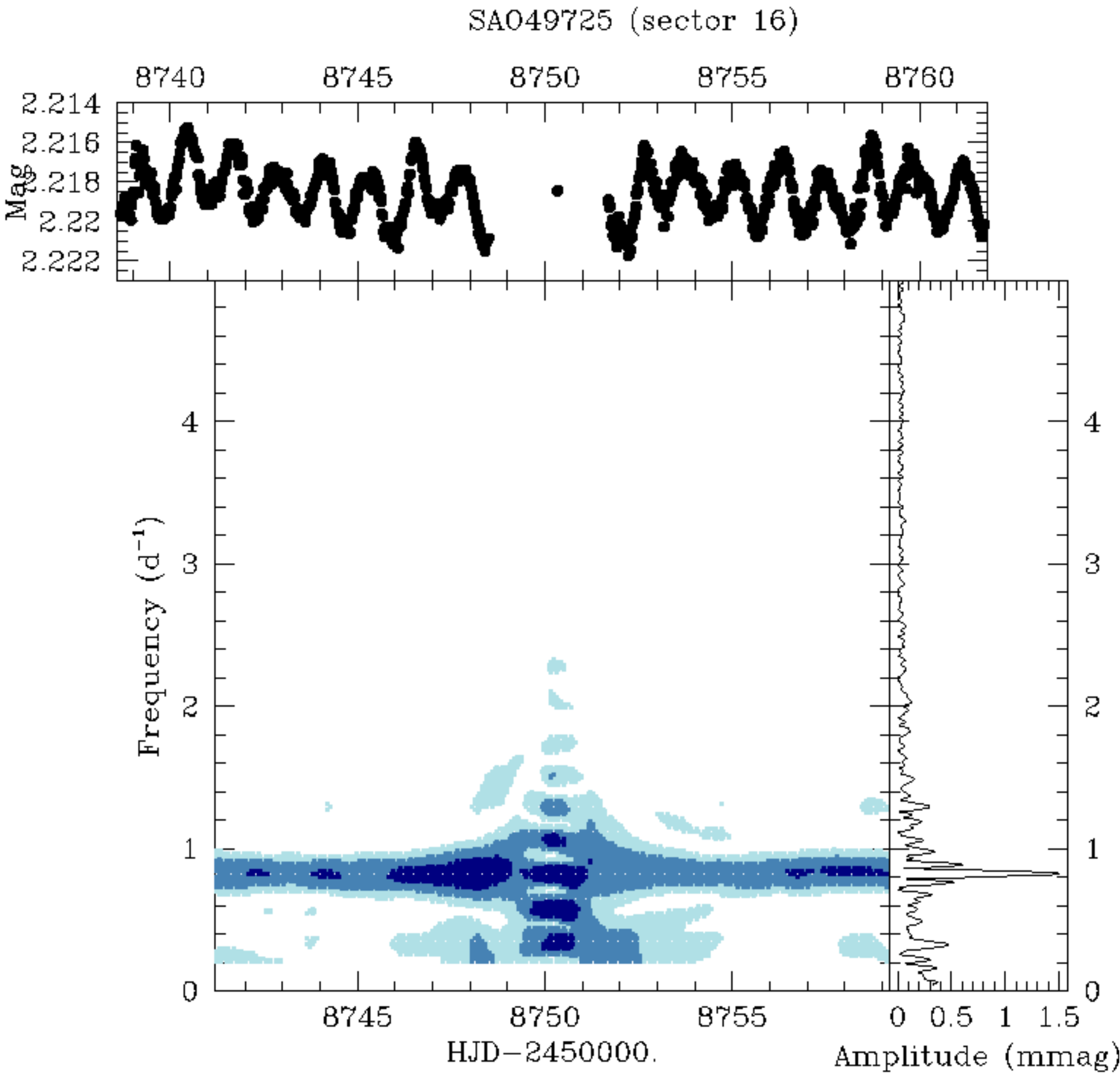}}
\resizebox{5.8cm}{!}{\includegraphics{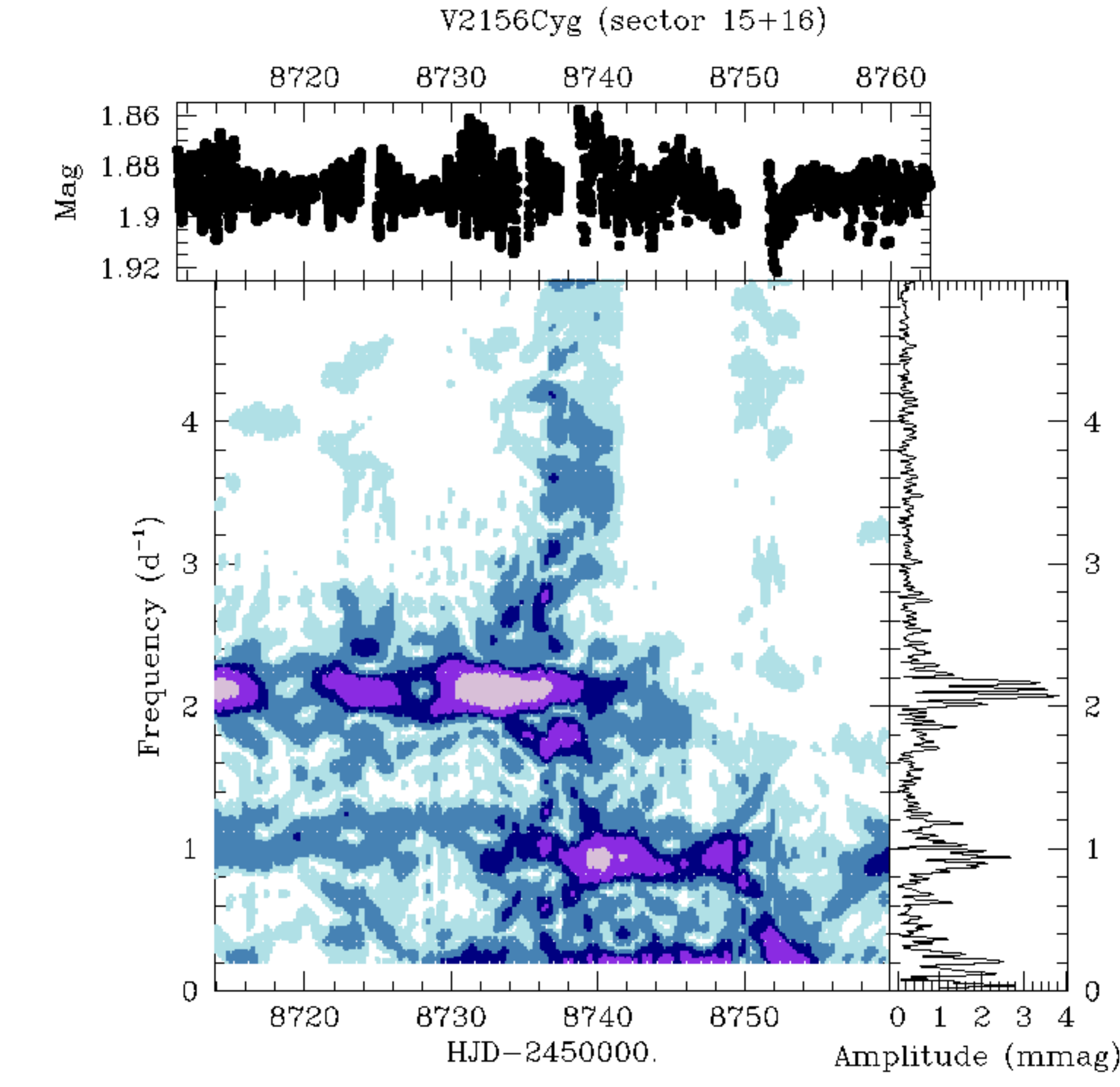}}
\resizebox{5.8cm}{!}{\includegraphics{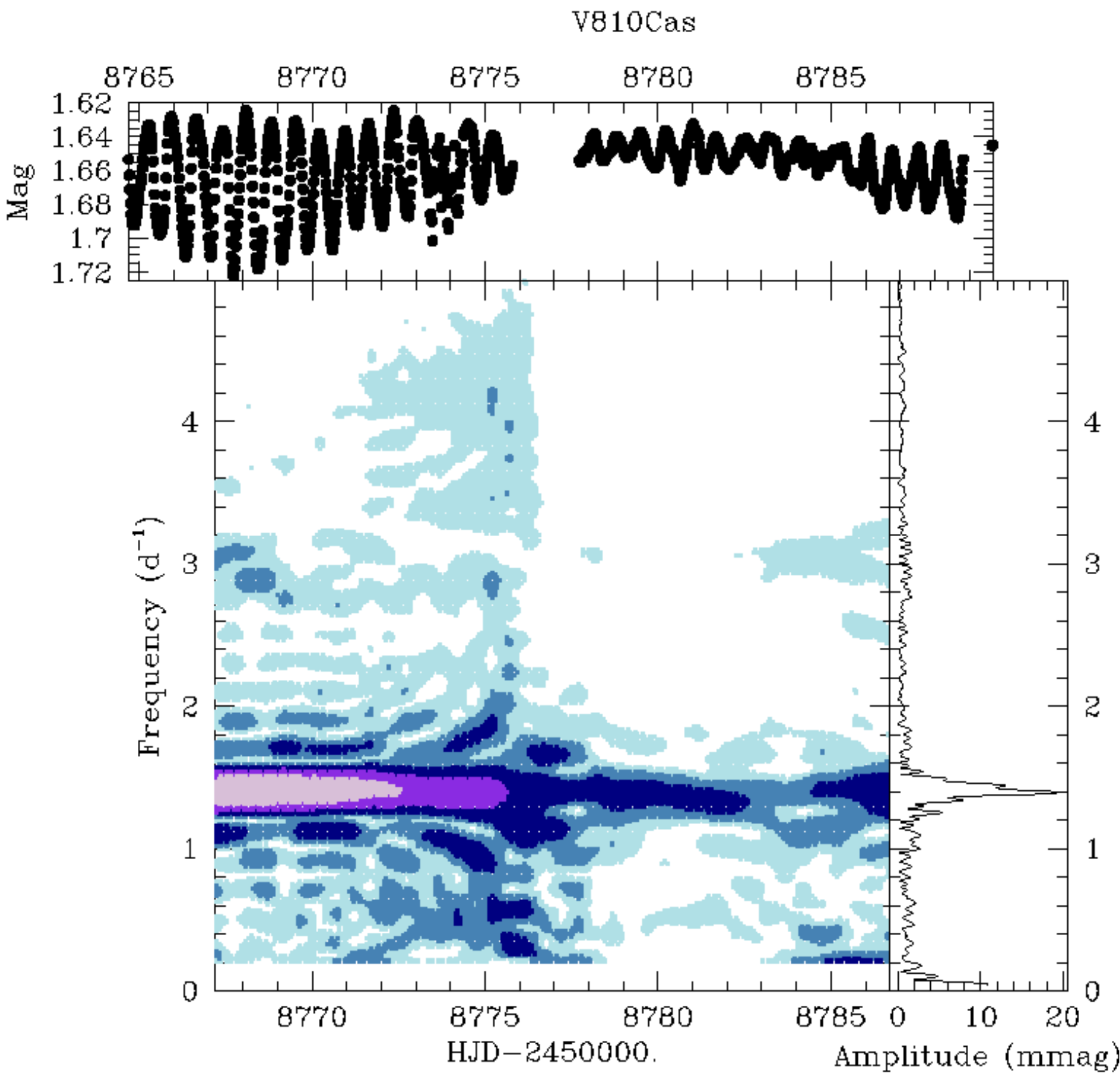}}
  \end{center}
  \caption{Continued.}
\end{figure*}

\subsection{Temporal evolution}
As mentioned at the end of Sect. 2, we also derived periodograms in sliding windows of 5\,d duration shifted by steps of 0.5\,d. These time-frequency diagrams  (Fig. \ref{evol}) enable to study the temporal evolution of the detected signals, although with a reduced frequency resolution (whatever the frequency, the natural peak widths are 0.04\,d$^{-1}$ for the full dataset of one sector, or 0.01--0.02\,d$^{-1}$ for multisector cases, but only 0.2\,d$^{-1}$ here). Only the periodograms of V782\,Cas, HD\,44458, HD\,45995, HD\,90563, and HD\,119682 appear relatively stable. Rather large changes are seen in all other cases, with two main categories. In the first category, the same frequencies (at the resolution limit) seem to be always present in the periodograms, but with changing amplitudes - varying oscillation amplitudes can also be spotted in lightcurves. This occurs in \gc, HD\,157832, V558\,Lyr, SAO\,49725, and V810\,Cas, as well as in HD\,110432 (especially for their main signals). Changes can be very localized (as for example in HD\,157832) or occurs more slowly. For example, the variations of SAO\,49725 were stronger in the first observing window (sector 14) than in the second one (sector 16) but the amplitude remains rather stable within a given window. An intermediate case is V810\,Cas, with its amplitude strenghtening covering one half of a sector. In TYC\,3681-695-1 and CQ\,Cir, the frequency groups also seem to be always present (with varying details, though). A second set of lightcurves display changes in both the frequency content and the amplitude of the signals. This seems to be the case for HD\,45314 (especially at low frequencies), V767\,Cen, and V2156\,Cyg. For the latter star, the signal near 2\,d$^{-1}$ clearly dominates during the first observing season (sector 15) whereas the second one (sector 16) rather shows low frequencies and signals near 1\,d$^{-1}$. 

In this context, it is interesting to search for links between the short- and long-term variabilities. For V767\,Cen, the short-term oscillations appear maximum towards the end of the observing window, more precisely at the maximum of a shallow brightness ``bump'' which itself follows a phase of slow decay. For this star, it does not seem that strong oscillations are triggering an outburst leading to a luminosity increase (as e.g. proposed for HD\,49330, \citealt{hua09}). In parallel, the short-term variations of TYC\,3681-695-1 have a larger amplitude during a brightness maximum but also during a luminosity increase occurring at the end of the observing window. 

\subsection{Variability in other datasets}

\begin{figure*}
\resizebox{5.5cm}{!}{\includegraphics{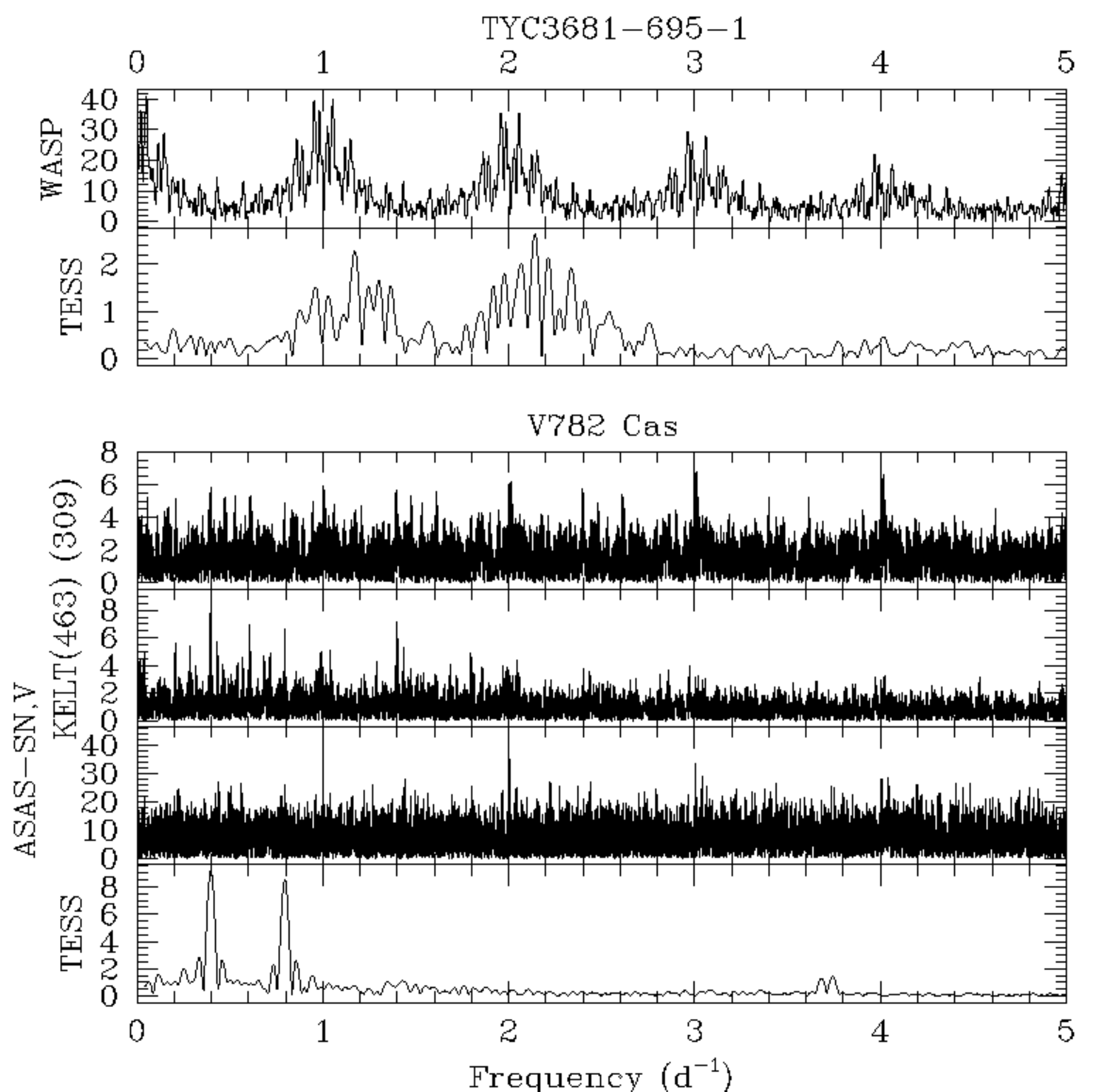}}
\resizebox{5.5cm}{!}{\includegraphics{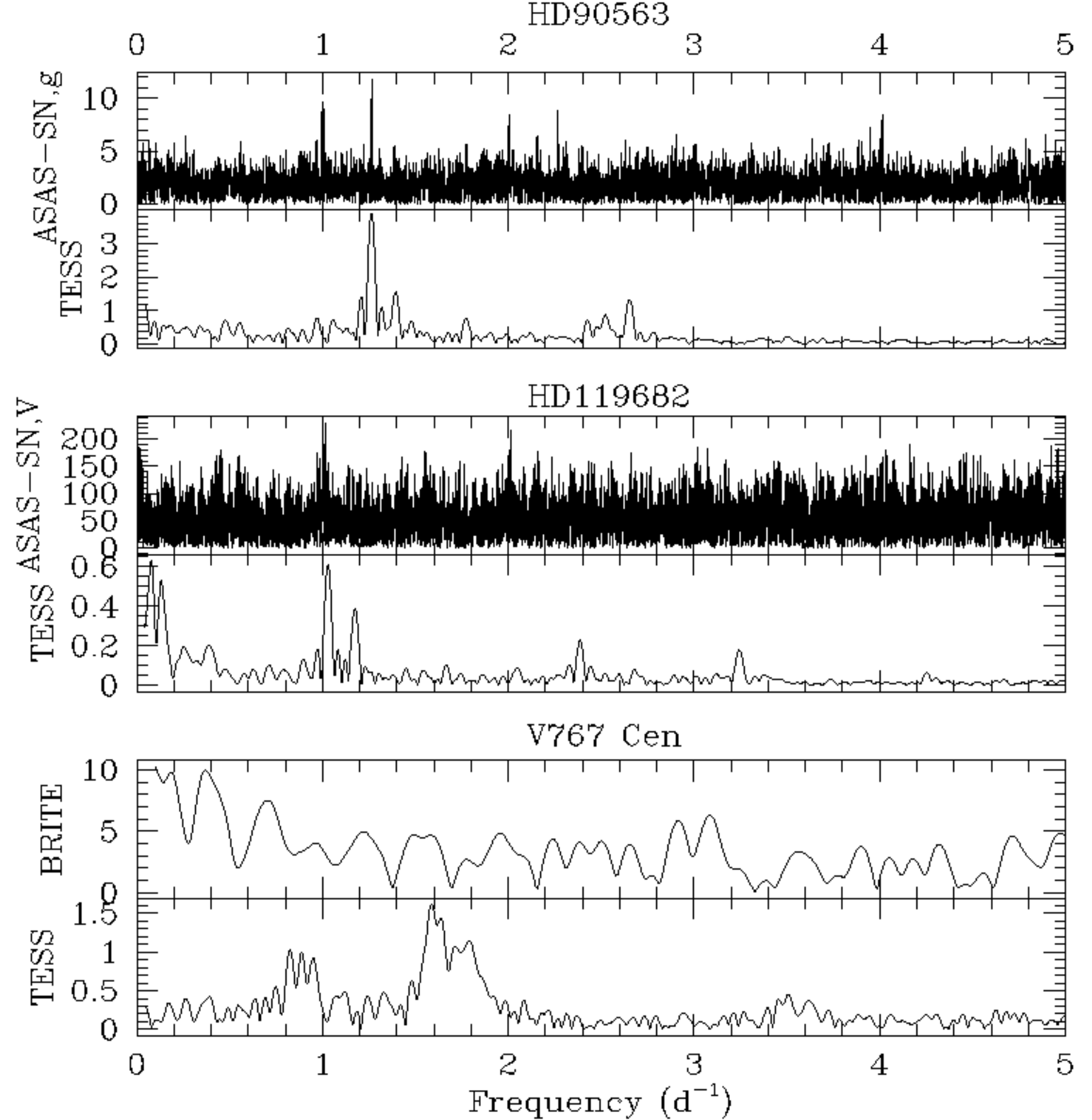}}
\resizebox{5.5cm}{!}{\includegraphics{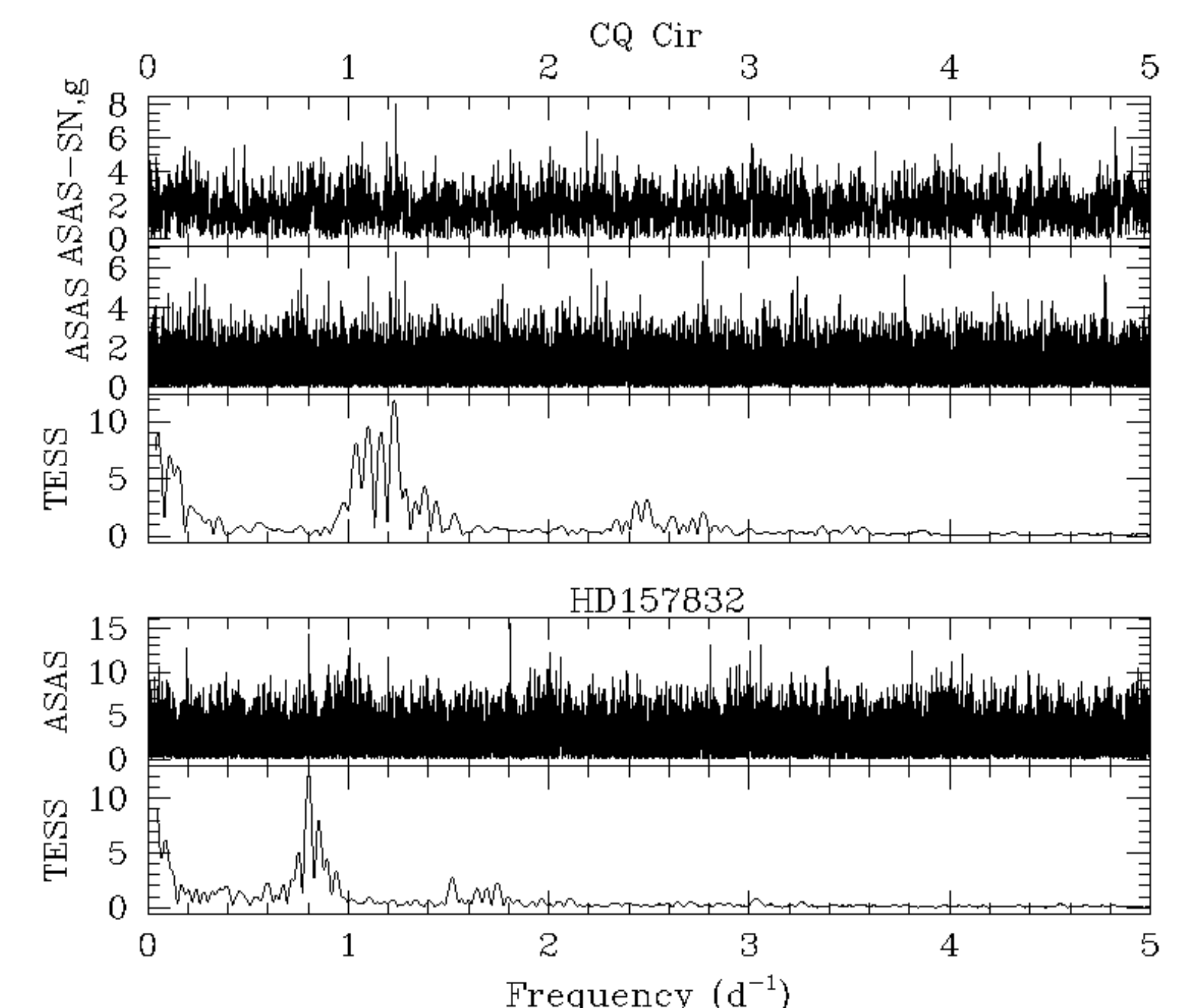}}
\resizebox{5.5cm}{!}{\includegraphics{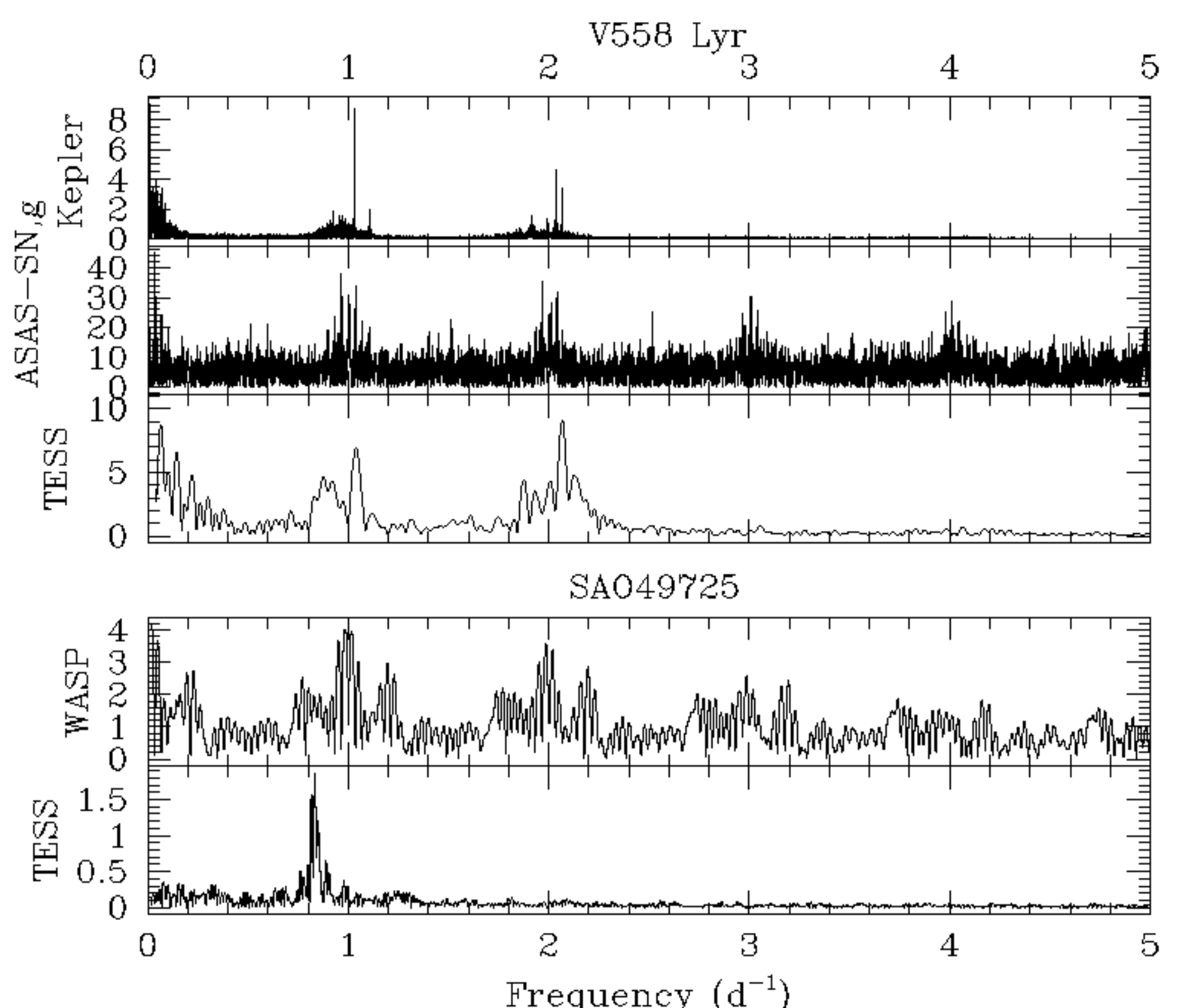}}
\resizebox{5.5cm}{!}{\includegraphics{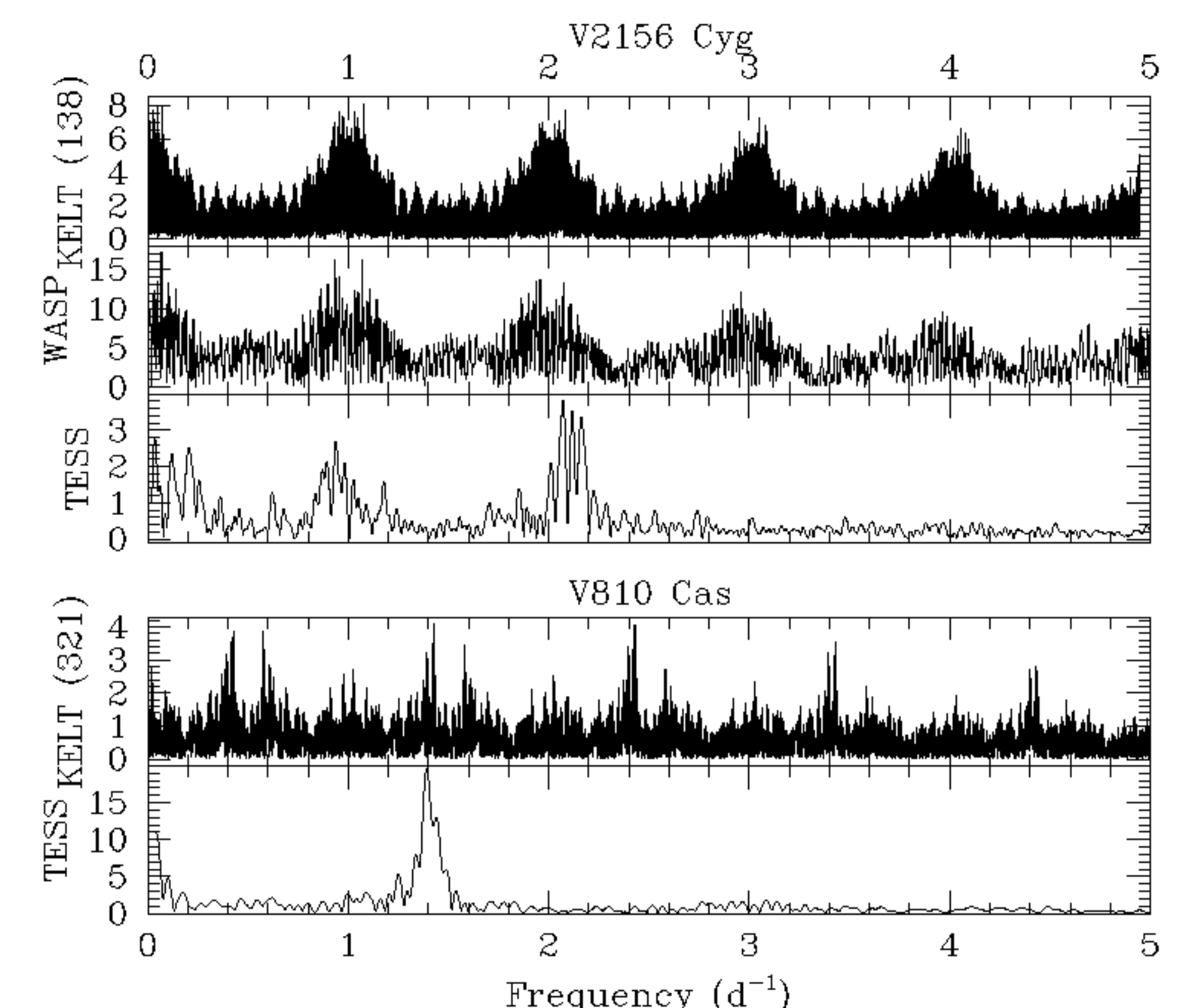}}
  \caption{Comparison between periodograms derived for four stars from \te\ and other observations; the ordinate provides sinusoid amplitudes in mmag. For {\it KELT}, the identification numbers from \citet{lab17} are quoted between parentheses. }
\label{compa}
\end{figure*}

The photometric variability of our targets has been previously studied, except for TYC\,3681-695-1, HD\,90563, HD\,119682, and SAO\,49725. In three other cases, the published information is very limited: HD\,45995 was reported in the {\it Hipparcos} catalog as photometrically variable, more precisely it had ``duplicity-induced variability'' \citep[see also][]{ols82}; HD\,45314 and CQ\,Cir were reported as photometrically variable in the {\it Hipparcos} catalog and by \citet{lef09}. No periodic signal was previously mentioned in the literature for those three stars, however. This work thus presents the first detailed photometric study for these targets. For the remaining stars, more information exists and it is mentioned in the dedicated items below, along with a comparison to our findings. To further facilitate this comparison, the last column of Table \ref{tab:var} lists the previously reported frequencies (also shown in Fig. \ref{ps} as vertical green lines), while Fig. \ref{compa} compares \te\ periodograms with those derived from published high-quality datasets from {\it Kepler} and {\it KELT} \citep{lab17,pop19}. 

In parallel, we searched for additional data of our targets in several extensive photometric databases: {\it SuperWASP}\footnote{https://exoplanetarchive.ipac.caltech.edu/cgi-bin/TblSearch/nph-tblSearchInit?app=ExoTbls\&config=superwasptimeseries}, {\it ASAS}\footnote{http://www.astrouw.edu.pl/asas/?page=aasc} \citep{poj97}, and {\it ASAS-SN}\footnote{https://asas-sn.osu.edu/} \citep{sha14}. For {\it SuperWASP}, as done for \te, the data points with errors larger than the mean of the errors plus their 1$\sigma$ dispersion were discarded. The derived periodograms are compared to the \te\ ones in Fig.\ref{compa}. For the latter two surveys, outliers and bad quality data were excluded but the noise often remained too large for detecting the subtle signals found in \te\ photometry. Therefore, Figure \ref{compa} only presents the favorable cases. A short (8\,d) high-cadence {\it BRITE}\footnote{http://brite.camk.edu.pl/pub/index.html} \citep{wei14} run was available for V767\,Cen and is here added for completeness, although it does not reveal any clear signal (Fig.\ref{compa}). 

Notes on individual stars:
\begin{itemize}
  \item \gc\ has been the target of many studies, including several on its photometric short-term optical variability. Using fifteen seasons of ground-based APT data, \citet{smi06} and \citet{hen12} detected a frequency of 0.82\,d$^{-1}$ with variable amplitude. The signal appeared lower in 2004--2011 compared to 1997--2003. \citet{bor20} confirmed these results using {\it SMEI} space-based data taken over the same years. In recent {\it BRITE} space-based observations, the strength of this signal seems to have further fainted, becoming undetectable. The dominant frequency now seems to be a signal at 2.48\,d$^{-1}$ frequency \citep{bor20}. The same authors also reported an additional signal at 1.25\,d$^{-1}$ as well as potential detections near 0.97 and 5.06\,d$^{-1}$. In the \te\ data, the main signal clearly is 2.48\,d$^{-1}$ and a peak at 5.054\,d$^{-1}$ also seems to be present. There is however no trace of the signals at 0.82, 0.97, and 1.25\,d$^{-1}$ - there are many small peaks in this region. Meanwhile, APT observations from the most recent seasons confirm the presence of the 2.48\,d$^{-1}$ and 5.06\,d$^{-1}$ signals; no other frequencies are visible in those data (Smith and Henry, in preparation).
  \item TYC\,3681-695-1 was also observed by {\it SuperWASP} (Fig. \ref{compa}). While some signal is detected in the periodogram, there are a strong daily aliasing and much larger amplitudes than in \te. The agreement with the frequency groups detected by \te\ near 1 and 2\,d$^{-1}$ is thus difficult to ascertain. 
  \item V782\,Cas was reported as ``unsolved variable'' in the {\it Hipparcos} catalog and in \citet{nic10}. Using {\it Hipparcos} data, \citet{lef09} derived a period of 47.621\,d and attributed an $\alpha$\,Cyg nature to the star. The \te\ lightcurve only covers 25\,d: this is more than half a 47.621\,d cycle and variations due to this cycle should thus be readily seen. However, no significant long-term modulation seems present for this star. More recently, {\it KELT} data indicated a period of 2.5131700\,d \citep{lab17}, which we confirm in \te\ data and possibly in {\it ASAS-SN} observations (Fig. \ref{compa}). These variations (Fig. \ref{lc}) present a strong similarity with eclipses, hence we further examine this possibility in the Appendix.
  \item HD\,44458 had been found to display quite large dispersions of its $V$ magnitude and $U-B$ colours in \citet{fei79} and was reported as ``unsolved variable'' in the {\it Hipparcos} catalog. Using {\it Hipparcos} data, \citet{hub98} derived a period of 0.536\,d which was confirmed by \citet{per02} using autocorrelation methods. This period is close to the strongest signal in the \te\ periodogram.
  \item HD\,90563 display the same main signal in \te\ and {\it ASAS-SN} data, though with a different amplitude (Fig. \ref{compa}).
  \item HD\,119682 has some signal in its {\it ASAS-SN} periodogram near the \te\ detection, but close to the daily alias and with a much larger amplitude, casting some doubt on the {\it ASAS-SN} detection.
  \item V767\,Cen has long been known to be variable, notably showing quite large dispersions of its $V$ magnitude and $B-V, U-B$ colour measurements \citep{cor66,fei75,fei79}. The presence of changes was confirmed by \citet{dac89}, \citet{doug94}, \citet{mou98}, and the {\it Hipparcos} data (the star appears as ``unsolved variable'' in the catalog). Further analysis of the {\it Hipparcos} data indicated outbursts recurring with a timescale of about 300\,d \citep{hub98} and a periodicity with $f$=0.23474\,d$^{-1}$ \citep{koe02}. The \te\ lightcurve indicates long-term variability, which could be in line with the recurrent outbursts, but shows no sign of the proposed period (Fig. \ref{ps}).
  \item CQ\,Cir shows hints of the principal peak in the main frequency group detected by \te\ in both {\it ASAS} and {\it ASAS-SN} data, though with varying amplitudes (Fig. \ref{compa}).
  \item HD\,157832 has been found to be variable long ago \citep{cou73,sta87}. It appears with a period of 1.10406$\pm$0.00006\,d in the {\it Hipparcos} catalog, which was confirmed by \citet{hub98}, \citet{per04}, and \citet{dub11}. In the \te\ data, the main period appears close but significantly longer than this value. However, with their overall scarcity, leading to strong aliasing, the {\it Hipparcos} data are not adapted for clear detections of such high frequencies. Besides, the peak in the \te\ periodogram appears somewhat broad, with a submaximum corresponding to the {\it Hipparcos} value (Fig. \ref{ps}). Finally, it may be noted that the {\it ASAS} data also confirm the \te\ results (Fig. \ref{compa}).
  \item V558\,Lyr was reported as ``unsolved variable'' in the {\it Hipparcos} catalog. \citet{mol05} detected two frequencies, 0.60972 and 0.62625\,d$^{-1}$, in {\it Hipparcos} photometry. More recently, \citet{pop19} found ``hump and spike'' features near 1 and 2\,d$^{-1}$ in {\it Kepler} data, without any signal near 0.6\,d$^{-1}$ (see their Fig. 11). We also fail to detect signals near 0.6\,d$^{-1}$ but \te\ observations, and possibly {\it ASAS-SN} data, confirm the {\it Kepler} signals (Fig. \ref{compa}).
  \item SAO\,49725 seems to display frequency groups in the {\it SuperWASP} periodogram, in particular signal exists at similar frequencies as in \te\ data, but there is not a dominant central peak. 
  \item V2156\,Cyg, reported as ``unsolved variable'' in the {\it Hipparcos} catalog, was found to have a period of 21.0346600\,d in {\it KELT} data \citep{lab17}, without outbursts \citep{lab18}. The star was observed in two consecutive sectors by \te, hence the data cover more than 50\,d. While there is no obvious trace of a 21\,d cycle in the \te\ lightcurve, the periodogram indicates the presence of a low-frequency signal, especially during the second observing window. However, its value is only marginally compatible with the {\it KELT} detection (strongest \te\ peak is at 28$\pm$3\,d in the combined periodogram). Figure \ref{compa} directly compares the {\it KELT}, {\it SuperWASP}, and \te\ periodograms: the presence of a strong daily alias renders the identification of the frequency groups detected by \te\ more difficult in the former two cases, but there seems to be an agreement between datasets.
  \item V810\,Cas, reported as ``unsolved variable'' in the {\it Hipparcos} catalog, displays slow variations according to \citet{lef09} but {\it KELT} data rather indicated a rather short periodicity \citep[2.3904100\,d, ][]{lab17}. The \te\ data reveal an even shorter period, of which the {\it KELT} detection is a daily alias: the absence of the \te\ signal at the {\it KELT} frequency and the strong daily aliasing in {\it KELT} periodograms are shown in Fig. \ref{compa}.
\end{itemize}

\section{Discussion}

In the previous section, we presented the analysis of the \te\ data of 15 \gc\ analogs. The targets display isolated frequencies or a few frequency groups, sometimes in harmonic relation and always appearing at low frequencies ($f<5\,{\rm d}^{-1}$ - only a few faint high-frequency signals are detected). The high-cadence, space-borne photometry of three \gc\ stars had already been studied (\gc, \citealt{bor20}; HD\,110432 and $\pi$\,Aqr, \citealt{naz20phot}), as well as the high-quality photometric datasets of several ``normal'' (i.e. non-\gc) Be stars \citep[e.g.][]{lab17,sem18,bal20}, allowing for a global comparison.

First, the global appearance of periodograms can be compared. Analyzing 57 Be stars observed by \te, \citet{bal20} reported that 40\% of the stars displayed a simple periodogram (one peak or two peaks in harmonic relation with little broadening or only some fine structure) while 30\% were more complex (frequency group or groups in harmonic relation). Amongst the 15 Be stars observed by {\it CoRoT} \citep{sem18}, about 30\% displayed strong frequency groups while a fifth of the sample had only a single peak dominating their periodogram. In the sample studied here, complemented by the previously studied cases of $\pi$\,Aqr and HD\,110432 \citep{naz20phot}, we found that $\sim$40\% of the stars display a relatively simple periodogram (i.e. with a dominant frequency) while a quarter shows strong frequency groups. Therefore, our \gc\ sample and the Be samples appear to be similar, especially considering the uncertainties due to small number statistics. Note that the separation between the two extreme cases mentioned here (single narrow vs broad groups) does not appear clear-cut, as intermediate cases (relatively strong peaks over fainter broad groups) are often seen in our sample. 

With the advent of high-precision space-borne photometry, it has also been found that a number of massive OB stars display elevated signal levels at low frequencies, aka ``red noise'', in their photometric times series \citep[e.g.][]{blo11,Tahina,rau19,bow19,Bow20}. Our \gc\ stars are no exception, showing that these stars fit the typical behaviour seen for stars of similar spectral types. The most popular scenarios to explain red noise variability are unresolved randomly excited internal gravity waves generated either in a subsurface convection zone or at the interface between the convective core and the radiative envelope \citep[see][and references therein]{AR15,Tahina,bow19,Bow20}. Concerning the latter, \citet{Rog13} showed that such internal gravity waves could transport significant amount of angular momentum to the stellar surface and thereby play a key role in the formation of Be decretion disks. 

The most frequent periods reported by \citet{lab17} for 510 Be stars observed by {\it KELT} without saturation are near 0.5 and 1.5\,d (i.e., frequencies near 0.7 and 2\,d$^{-1}$), in line with our results. Similar conclusions are reached by looking at the periodograms shown by \citet{sem18} and \citet{bal20}. It may be further noted that three quarters of our 15 targets display coherent signals with $P<2$\,d, while at least 28\% of early-type Be stars of the {\it KELT} sample do. Besides small number statistics, the lower percentages of such detections in the {\it KELT} sample may probably be explained: using space facilities allows for detecting fainter and higher frequency signals (because of improved sensitiviy and denser sampling) compared to ground-based telescopes. 

When interpreted in terms of pulsational activity, the coherent signals at low frequencies could correspond to g-modes as in SPB stars while the broad frequency groups could represent dense populations of low-order g-modes \citep{cam08}. In this context, the limitations of the \te\ datasets should be taken into account. Indeed, the typical month-long observation of one sector leads to a natural peak width of $\sim$0.04\,d$^{-1}$. Furthermore, the sampling (in particular the gap splitting the sector observations in two separate observing windows of about 2 weeks duration) creates aliases at $f_{peak}\pm0.06...\,{\rm d}^{-1}$ (see Fig. \ref{sw} for details). All this renders the identification of real subpeaks difficult as typical mode separations of 0.01\,d$^{-1}$ are often found, see e.g. the zooms on groups shown by \citet{sem18}. The presence of real flanks can however be hinted at, as prewhitening by the peak frequency often leads to remaining wings in the periodograms (aliases will on the contrary be cleaned along with the main frequency). Longer observations, coupled to an intense spectroscopic monitoring, would certainly help confirming the presence of modes and identifying them.

Another possibility should also be considered: r-modes \citep{sai18}. These modes are global Rossby waves which appear in rotating stars. Because of their origin, such modes should be especially frequent in rapidly rotating stars such as Be stars \citep{sai18}. In a periodogram, they should form a frequency group appearing below the rotation frequency, hence their appearance of a ``hump'' dominated by a strong peak at its high-frequency limit (explaining the ``hump and spike'' name chosen by \citealt{sai18}). In our targets, this is clearly the case of V558\,Lyr, as was already remarked by \citet{pop19}. HD\,45995 and CQ\,Cir could be two additional cases, but this needs to be confirmed with data taken on longer timebases, to improve the frequency resolution.

In addition to (or in place of) pulsations, rotational modulation may also arise if starspots are present on the stars. Since spots create not fully coherent variability, this could lead to frequency groups \citep{bal20}, allowing for an alternative interpretation of these features.

The last possibility to explain (some of) the flux variations relates to the circumstellar disk \citep[e.g.][]{riv16}. Indeed, low frequencies, close to stellar rotation, may arise from material orbiting in the inner parts of the disk. Frequency groups could then arise from circumstellar aperiodic variability superimposed on more regular features. Reprocessing of (varying) stellar illumination may also lead to apparent disk variability. Finally, in cataclysmic variables such as Z\,Cam, large but intermittent brightness oscillations have been recorded. They may be somewhat reminiscent of the sudden change in oscillation amplitude seen here for e.g. V810\,Cas, though with much smaller flux variations and on much shorter timescales. Such changes are explained by a thermal-viscous instability linked to hydrogen ionisation variations in the disk \citep[hysteresis between a bright, hot, high-viscosity state and a faint, cool, low-viscosity state, ][]{ham19}. This model was extended to accretion disks in X-ray binaries and FU\,Ori-type stars, but it remains to be seen whether decretion disks around Be stars could also undergo a similar instability.

\begin{figure}
  \begin{center}
\resizebox{8cm}{!}{\includegraphics{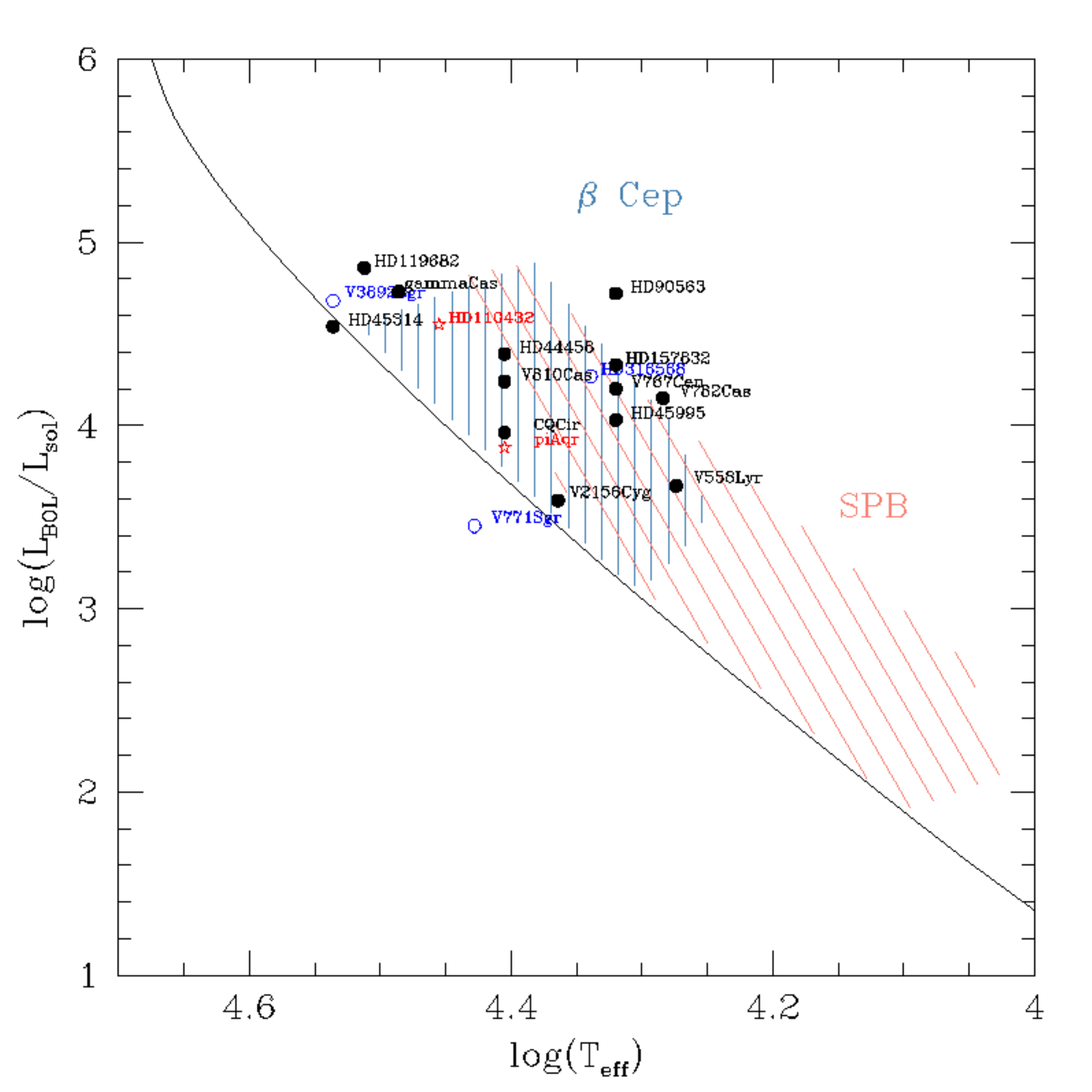}}
  \end{center}
  \caption{HR diagram of our targets (black dots), three previously studied \gc\ stars (red stars), and three other, unobserved \gc\ stars (blue circles). The temperature and luminosity from \citet{naz18} and \citet{naz20x} are used. The ZAMS from Geneva stellar evolution models (https://www.unige.ch/sciences/astro/evolution/fr/base-de-donnees/syclist/) for solar abundance and no rotation is indicated, as well as the instability zones of $\beta$\,Cep and SPB stars from \citet{mig07}. }
\label{hr}
\end{figure}

While low-frequency peaks, isolated or in groups, are detected, signals at higher frequencies ($f>5\,{\rm d}^{-1}$) are also recorded in some stars. However, they seldom appear alone or strong, i.e. dominating the frequency content or even simply with amplitudes comparable to the low-frequency signals. In \citet{bal20}, only $\sim$10\% of the stars displayed such strong high-frequency peaks, whereas one fifth of the 15 stars analyzed by \citet{sem18} do so. For \gc\ analogs, only two stars ($\pi$\,Aqr and HD\,110432) harbour such strong signals (which makes a fraction around 10\%). A systematic search for signals with smaller amplitudes should certainly be undertaken, but it can already be said that half of the sample of \citet{sem18} and half of our sample display peaks, faint or (relatively) strong, at these frequencies. Again, there thus seems to be a good agreement between samples. Such coherent high-frequency signals are often interpreted as p-modes and it is therefore useful to see where the targets appear in the HR diagram. When the stellar parameters from \citet{naz18} and \citet{naz20x} are used, the \gc\ analogs are grouped in the HR diagram. Some stars appear outside of the SPB locus, though they vary at low frequencies, but all appear within the $\beta$\,Cep locus (Fig. \ref{hr}), hence could potentially pulsate at high frequencies\footnote{On the contrary, in the sample of \citet{sem18}, some stars with high-frequency signals appear outside of the $\beta$\,Cep locus.}. While there remain uncertainties on the exact positions of these Be stars in the HR diagram, because of the fast rotation and unknown viewing angle which could bias the luminosity and temperature evaluations, the presence of high-frequency signals, strong or faint, in only some objects is not readily explained. The question probably needs to be assessed through e.g. detailed asteroseismic modelling of these stars.

In parallel, combinations of frequencies have been reported several times in the periodograms of Be stars (see e.g. \citealt{sem18}, in which 40\% of targets present it). Our \te\ sample is no exception, with about one third of stars potentially exhibiting such combinations (Table \ref{tab:var}). HD\,44458 has a peak at 0.816\,d$^{-1}$ which could correspond to the difference between the 1.084 and 1.904\,d$^{-1}$ signals. For HD\,45314, the broad structure peaking at 0.288\,d$^{-1}$ could be compatible with the difference between the 1.184 and 1.460\,d$^{-1}$ signals. In HD\,90563, the difference between signals at 1.264 and 2.656\,d$^{-1}$ is close to the 1.396\,d$^{-1}$ frequency. For HD\,119682, the 0.140\,d$^{-1}$ peak nicely fits the difference between the 1.032 and 1.176\,d$^{-1}$ signals. And for SAO\,49725, the difference between the two faint high-frequency signals at 7.359 and 8.194\,d$^{-1}$ appears close to the frequency of the main signal at 0.833\,d$^{-1}$.

Finally, most, if not all, Be stars present long-term variations, generally interpreted in terms of disk building/dissipation. About half of the early Be stars observed by {\it KELT} displayed outbursts and 45\% showed long-term trends \citep{lab17}. In the {\it CoRoT} sample, \citet{sem18} reported outbursts for $\sim$40\% of the stars. For \gc\ analogs \citep[and this work]{naz20phot}, four cases of strong (a few 0.01\,mag) brightness variations are detected, which represents a quarter of the sample. This is a lower fraction than for {\it KELT} but the coverage of \te\ data (typically a month) is short compared to long ground-based campaigns such as {\it KELT}, which limits the possibility to observe outbursts or long-term signals.

In this context, it is important to note that the properties of the short-term oscillations do not remain constant over time, for the \gc\ stars as for other Be stars \citep{bal20}. Changes are particularly interesting to note during outbursts. The {\it CoRoT} cases displayed larger oscillations during outbursts (corresponding to decreasing or fading brightnesses). Generally, the amplitudes of the main frequencies increased during these events, with only fainter signals sometimes doing the contrary. In \citet{hua09} and \citet{bal20}, the outbursting HD\,49330 showed a decrease in the amplitude of its high-frequency signals while low-frequency groups appeared. In $\pi$\,Aqr, some high-frequency signals decreased while other increased (or appeared) during outbursts while HD\,110432 showed larger amplitudes for both low- and high-frequency signals when the star was brighter \citep{naz20phot}. In the two cases reported in the previous sections, an increase in oscillation amplitude was detected at the maximum of an outburst and afterwards, not before it. In the small set of stars whose oscillations were examined at and near outbursting events, there thus seems to be no general rule, as e.g. stronger pulsations just before Be outbursts. Besides, peak changes of similar amplitudes can be noted outside outburst events (see e.g. HD\,157832, V2156\,Cyg, V810\,Cas). Definitely, more stars should be followed in detail before drawing conclusions on the triggering effects associated to outbursts and pulsation changes.

\section{Summary and conclusions}
This paper reports on the analysis of \te\ lightcurves of 15 \gc\ analogs, complemented by {\it ASAS, ASAS-SN, BRITE, SuperWASP} archival data. It enlarges a previous study analyzing high-cadence photometric data of two bright \gc\ stars. This work allows us to constrain the short-term variability properties for two thirds of the \gc\ sample. Studying such a significant fraction provides clear insights on the whole class.

The periodograms of these 15 \gc\ stars present signals at low frequencies ($f<5\,{\rm d}^{-1}$), with only faint (though significant) signals above that limit. The low-frequency signals consist in broad frequency groups, a narrow dominant peak, or a combination of features (a mix of isolated coherent signals and/or frequency groups). Red noise also appears to be frequent, as in other OB stars. Amplitudes of signals in periodograms often change over time, but the frequency content may also be altered. All these detected features are also observed in ``normal'' Be stars, suggesting an absence of defining \gc\ characteristics in the optical photometric variability. Comparing the two recorded long-term variations of our sample with published cases, there does not seem to be a general rule (yet?) relating outburst occurence and changing oscillation amplitudes in Be stars. 

\section*{Acknowledgements}
The authors thank L. Eyer for valuable discussions on {\it Gaia} data, and J. Labadie-Bartz and J. Nichols for helping locate their published data. Y.N. and G.R. acknowledge support from the Fonds National de la Recherche Scientifique (Belgium), the European Space Agency (ESA) and the Belgian Federal Science Policy Office (BELSPO) in the framework of the PRODEX Programme (contract HERMeS). AP acknowledges support from NCN grant no. 2016/21/B/ST9/01126. ADS and CDS were used for preparing this document. 

\section*{Data availability}
The data underlying this article are available in repositories: \te\ data are all available from the MAST archives, ASAS, ASAS-SN, SuperWASP, and BRITE data are also archived and available online - see footnotes 1,2,4,5,6,7 for details.

\appendix

\section{V782\,Cas}
The light curve of V782\,Cas is reminiscent of that of an eclipsing binary in an overcontact configuration. Assuming the Be star to be one of the components of that eclipsing binary leads however to a major contradiction. Indeed, TIGRE spectroscopic observations taken on 1 November, 18 November and 2 December 2019 (i.e.\ immediately before, at the middle of, or immediately after the \te\ campaign) clearly indicate a strong H$\alpha$ emission line with an equivalent width between $-24.1$ and $-24.5$\,\AA. The separation of the blue and red peaks varied between 2.95\,\AA\ and 3.28\,\AA, corresponding to a separation in radial velocity of 135 -- 150\,km\,s$^{-1}$. Adopting $v\,\sin{i} = 188$\,km\,s$^{-1}$ \citep[see][and references therein]{naz18}, relation (2) of \citet{Zamanov} for a Keplerian disk leads to $R_{\rm disk} \simeq 6.3 -7.8$\,R$_*$. Hence, the Be star is surrounded by a wide decretion disk which is not compatible with the Be star being in an overcontact configuration. With $\log{L_{\rm bol}} = 4.15$ and $T_{\rm eff} = 19\,230$\,K \citep{naz18}, we can estimate a stellar radius of 10.7\,R$_{\odot}$. Assuming a typical mass of 10\,M$_{\odot}$ for the B2.5\,IIIe star, we find that the semi-major axis of a binary system with a period of 2.51\,days (corresponding to the detected frequency of $\sim0.4$\,d$^{-1}$) amounts to $16.8\,(1+q)^{1/3}$\,R$_{\odot}$ where $q$ is the mass ratio between the secondary and the B2.5\,IIIe primary. Even at the hard limit of $q \rightarrow 0.0$, the radius of the Roche lobe remains below 13\,R$_{\odot}$ which is totally inconsistent with the presence of a large circumstellar disk around the primary, unless the Be decretion disk would be circumbinary, which would make this system a very unusual case.

The most likely alternative then appears to be either a fortuitous line-of-sight alignment between the Be star and a short-period eclipsing binary or a genuine triple system consisting of the B2.5\,IIIe star physically bound to the eclipsing binary. Concerning the first possibility, we have checked the {\it Gaia}-DR2 database for contaminating sources within the 1\,arcmin extraction radius of the \te\ photometry. V782\,Cas has a $G$ magnitude of 7.48. The next brightest object has $G = 11.7$, followed by stars with $G \ge 12.9$. Since the maximum depth of the primary eclipses amounts to 0.02\,mag, whereas the secondary eclipses reach a depth of $\sim 0.01$\,mag, the eclipsing binary would contribute at least $\sim 3$\% of the total light in the \te\ photometry. If the {\it Gaia} and \te\ photometric bands were identical, this result would imply that the line-of-sight object must be of magnitude $G = 11.3$ or brighter. Taking into account the fact that the \te\ band peaks at longer wavelengths than the {\it Gaia} band, the object at $G = 11.7$ ({\it Gaia}-ID 515475966211949056) would be the most plausible candidate among the resolved neighbours.

\begin{figure}
  \begin{center}
\resizebox{8cm}{!}{\includegraphics{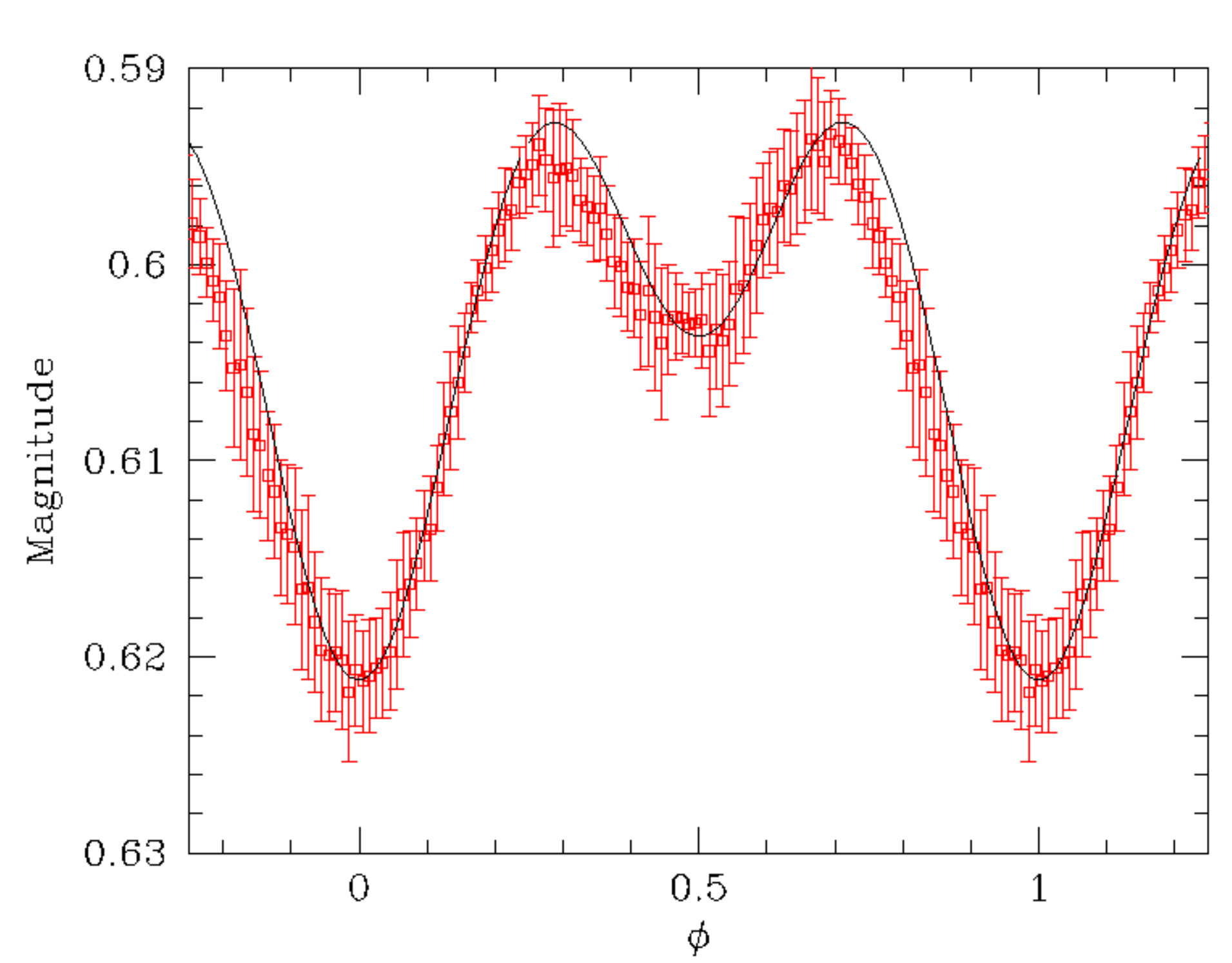}}
 \end{center}
  \caption{Binned \te\ lightcurve (red open rectangles) and best-fit eclipse solution (see text for details).  }
\label{fitnightfall}
\end{figure}

To investigate this possibility, we folded the \te\ data using $T_0=2\,458\,790.7$ and $P=2.51256$\,d and then built a lightcurve consisting of 100 normal points computed as the mean magnitude per 0.01 phase bin. We took the dispersion of the data points within a phase bin as our estimate of the uncertainty. We then used the {\tt Nightfall} code \citep{Wichmann} to fit the light curve consisting of the normal points with a binary light curve model accounting for the presence of a third light and for reflection between the stars. Since the light curve fit is usually not very sensitive to the mass ratio, we fixed this parameter to unity. Lacking any information on the temperatures of the components of the eclipsing binary, we set the primary temperature to 6000\,K or 10\,000\,K, and kept the secondary temperature as a fitting parameter. Setting the values of the third light (i.e., the Be star) contribution to values of 95\% and assuming Roche-lobe filling factors below unity did not allow us to achieve an acceptable fit of the light curve. We thus tested overcontact configurations (i.e.\ values of the Roche-lobe filling factor exceeding one). This improved the quality of the fit though there were still systematic deviations between the observational light curve and the model, and the best-fit temperature of the secondary star became unrealistically low ($< 100$\,K).  We then allowed the code to adjust the value of the third light. This led to a significant improvement of the global adjustment, whilst simultaneously leading to lower values of the percentage of third light. For instance, the solution shown in Fig.\,\ref{fitnightfall} corresponds to Roche lobe filling factors of 1.15 for both stars, an inclination of $27.9^{\circ}$, primary and secondary temperatures of respectively 10\,000 and 2600\,K, and a third light contribution of 75\%. This low value for the third light would imply that the eclipsing binary is not the neighbour with $G=11.7$, rather suggesting a system unresolved from (hence possibly bound to) the Be star. We caution however that given the lack of constraints on the properties of the companions forming the eclipsing binary, there is certainly no unique solution. Our purpose here was only to check whether or not such an eclipsing binary scenario is plausible.

\bsp	
\label{lastpage}
\end{document}